\pdfoutput=1
\RequirePackage{ifpdf}
\ifpdf %We are running pdfTeX in pdf mode
\documentclass[pdftex]{sigma}
\else
\documentclass{sigma}
\fi

\numberwithin{equation}{section}

\usepackage{MnSymbol}

\begin{document}

\newcommand{\arXivNumber}{1312.6073}

\allowdisplaybreaks

\renewcommand{\PaperNumber}{092}

\FirstPageHeading

\ShortArticleName{Beables/Observables in Classical and Quantum Gravity}

\ArticleName{Beables/Observables in Classical\\
and Quantum Gravity}

\Author{Edward ANDERSON}

\AuthorNameForHeading{E.~Anderson}

\Address{DAMTP, Centre for Mathematical Sciences, Wilberforce Road, Cambridge CB3 0WA, UK}
\Email{\href{mailto:ea212@cam.ac.uk}{ea212@cam.ac.uk}}

\ArticleDates{Received December 20, 2013, in f\/inal form August 18, 2014; Published online August 29, 2014}

\Abstract{Observables `are observed' whereas beables just `are'.
This gives beables more scope in the cosmological and quantum domains.
Both observables and beables are entities that form `brackets' with `the constraints' that are `equal to' zero.
We explain how depending on circumstances, these could be, e.g., Poisson, Dirac, commutator, histories,
Schouten--Nijenhuis, double or Nambu brackets, f\/irst-class, gauge, linear or ef\/fective constraints, and strong, weak or
weak-ef\/fective equalities.
The Dirac--Bergmann distinction in notions of gauge leads to further notions of observables or beables, and is tied to
a~number of dif\/feomorphism-specif\/ic subtleties.
Thus we cover a~wide range of notions of observables or beables that occur in classical and quantum gravitational
theories: Dirac, Kucha\v{r}, ef\/fective, Bergmann, histories, multisymplectic, master, Nambu and bi-.
Indeed this review covers a~representatively wide range of such theories: general relativity, loop quantum gravity,
histories theory, supergravity and M-theory.}


\Classification{83C05; 83C45; 83D05; 70H45; 81S05}

\section{Introduction}%\label{Intro}

This review covers a~topic~-- observables and beables~-- which spans classical dynamics and quantum mechanics, with the
canonical perspective of each of quantum cosmology and quantum gravity particularly in mind.

Observables/beables~\cite{Bergmann61, BK72, DiracObs, Dirac, Dittrich1,
Dittrich2, G-M-H,HT92,I93, Kuchar92, Kuchar93, PSS10,Rovellibook, Thiemannbook, Tambornino} are often considered to be objects whose `brackets' with `the
constraints' are `equal to' zero:
\begin{gather}
|\boldsymbol{[}{\cal C}_{\text{\sffamily{C}}},
B_{\text{\sffamily{B}}}\boldsymbol{]}|
\, \text{`='}\,  0.
\label{basic-obs}
\end{gather}
Here ${\cal C}_{\text{\sffamily{C}}}$ denotes the constraints and $B_{\text{\sffamily{B}}}$
denotes the beables; $\text{\sffamily C}$ and $\text{\sffamily B}$ index for now general sets of each of these.
$|\boldsymbol{[}~,~\boldsymbol{]}|$ is usually a~{\it Lie bracket} such as a~Poisson bracket in
classical dynamics or a~quantum commutator.
As a~Lie bracket, it obeys the {\it Jacobi identity}
\begin{gather*}
|\boldsymbol{[} X , |\boldsymbol{[} Y , Z \boldsymbol{]}|  \boldsymbol{]}| + \text{cycles} = 0.
\end{gather*}
However, there are a~number of dif\/ferent possibilities for which brackets, which constraints and even which notion of
equality can be involved.
Thus we will f\/irst need to discuss each of these more primary entities (Sections~\ref{1.1}--\ref{1.7})
with additionally some types of constraint having ties to notions of gauge.
Additionally, there are notions of gauge {\it not} tied to constraints which furnish a~further conception of
observables/beables along the lines of Bergmann~\cite{Bergmann61, BK72}.
After this, we can return to considering the more composite notions that are observables and beables, in Sections~\ref{1.8}--\ref{1.10}.

We do f\/irst consider the distinction between {\it observables} and {\it beables}.
This began with Bell~\cite{Bell75}, and is the dif\/ference between entities being observed and entities simply being.
The circumstances under which observables occur are then a~subset of those in which beables do, in the sense that `being
observed' is a~subset of `being'.
Moreover, from a~beables perspective, def\/ining what `observing' is is unnecessary, so conceptualizing in terms of
beables is a~freeing from having to def\/ine this.

\looseness=-1
Two contexts in which beables are relevant are 1) whole-universe or closed-system modelling, and 2) at the quantum level.
Bell pointed out that 1) is already an issue at the classical level~\cite{Bell}.
This is due to observers living within such a~universe rather than af\/fording a~`God's eye' view from outside.
(Also observers did not exist in the early universe.)   On the other hand, 2) has more widespread relevance due to the
connection between the notion of observation and the quantum measurement problem~\cite{Measurement1, Measurement2}; Bell
furthermore extended the notion of beables to QFT in~\cite{Bell87}.

Along lines e.g.\ recently argued for by Kent~\cite{Kent}, beables are furthermore an appropriate concept for a~number of types of {\it
realist interpretation} of QM.
(This is as opposed to e.g.\ {\it instrumentalist interpretation}; see e.g.~\cite{IBook} for an introduction to dif\/ferent types of interpretation of
QM.)   The Bohmian approach is one branch of realist interpretations in which the name and concept of `beables' is
widely used~\cite{Bohm8,Bell87,Bohm2,Bohm4, Bohm9, Bohm3, Bohm6,Bohm5, Bohm7, Bohm10, Bohm1}.
Moreover, histories approaches (see~\cite{GMH, Hartle95, IL}, or Appendix~\ref{PoT} for a~summary) can also be thought of in
terms of beables.
The beables concept additionally comes hand-in-hand with QM wavefunction collapse due to decoherence~\cite{Giu}~by
natural phenomena as opposed to by observation.
I.e.\
in the Universe, processes such as dust grains being decohered by CMB photons~\cite{JZ} are more typical than processes
specif\/ically involving observers making quantum measurements.
Finally, beables are also appropriate in the {\it contextual realist} interpretation of QM by Isham and
Doering~\cite{ID}\footnote{This   approach is based on multi-valued context-dependent truth valuations via use of Topos
Theory to reinterpret the foundations of QM.
As such this approach is given here as motivation for realist approaches, but further details of it lie beyond the scope
of this review.}.   Among these realist approaches, histories and decoherence play further role in this review.

As a~third context combining the previous two, quantum cosmology is substantially distinct from QM.
The measurement problem is further aggravated in this setting, for which the usual Copenhagen interpretation of QM can
no longer apply.
Quantum cosmology has its own distinct conceptualization of histories and decoherence~\cite{H03, Hartle95, Giu}.
Yet the concept of beables continues to be appropriate in quantum cosmology.

Henceforth I use `beables' unless the situation specif\/ically requires use of the word `obser\-vables'.

\subsection{Outline of constrained dynamics and various kinds of constraints}\label{1.1}

Denote the generalized conf\/igurations of a~physical system by $Q^{\text{\sffamily{A}}}$.\footnote{In this
review,   sans-serif capital letters are used as generalized indices, lower-case Latin letters are used for spatial
indices, and lower-case Greek letters for spacetime ones.
Primed and unprimed indices index the same objects throughout this review.
Following~\cite{I93, Kuchar92}, I use (~) for functions, [~] for functionals, and (~;~] for mixed function-functionals.
This leaves \{~\} without commas for actual brackets.
I then use bold font to clearly distinguish Poisson brackets $\boldsymbol{\{}~,~\boldsymbol{\}}$ and
other brackets playing analogous roles in def\/ining notions of beables.}     E.g.\
particle positions in mechanics, f\/ield values in f\/ield theory, or spatial 3-metrics $h_{ij}$ on a~f\/ixed topological
manifold~$\Sigma$ in the geometrodynamical formulation of general relativity (GR): Wheeler's~\cite{Battelle} formulation
of GR as evolving spatial 3-geometries.
The space of possible values that the $Q^{\text{\sffamily{A}}}$ can take is the {\it configuration space}
$\mathfrak{q}$~\cite{Lanczos}, e.g.\
$\mathbb{R}^{Nd}$ for~$N$ particles in dimension~$d$, or the space Riem($\Sigma$) of $h_{ij}$'s for geometrodynamics.
A~given (for now f\/inite) classical physical system's equations of motion can be taken to follow from the Lagrangian
$\text{\sffamily L}(Q^{\text{\sffamily{A}}}, \dot{Q}^{\text{\sffamily{A}}})$.\footnote{Here  \looseness=-1
$\dot{~}$
is $\partial/\partial t$ for~$t$ a~notion of time for one's theory, which includes in some cases
$\partial/\partial\lambda$ for~$\lambda$ a~meaningless label time.
This formula and the rest in this section are for f\/inite models such as mechanics or minisuperspace (homogeneous GR),
but have well-known extensions to f\/ield theories (including GR and alternative theories of gravity).}     The
$Q^{\text{\sffamily{A}}}$ then have conjugate momenta
\begin{gather}
P_{\text{\sffamily{A}}}:= \partial \text{\sffamily L}/\partial\dot{Q}^{\text{\sffamily{A}}}.
\label{mom-vel}
\end{gather}
One can additionally pass (via a~so-called {\it Legendre transformation}) from $Q^{\text{\sffamily{A}}}$ and
$\dot{Q}^{\text{\sffamily{A}}}$ variables and a~Lagrangian function of these, $\text{\sffamily
L}(Q^{\text{\sffamily{A}}}, \dot{Q}^{\text{\sffamily{A}}})$, to $Q^{\text{\sffamily{A}}}$ and $P_{\text{\sffamily{A}}}$ variables and a~Hamiltonian function of these, $\text{\sffamily
H}(Q^{\text{\sffamily{A}}}, P_{\text{\sffamily{A}}})$.
{\it Phase space} is the space of both the $Q^{\text{\sffamily{A}}}$ and the $P_{\text{\sffamily{A}}}$ as equipped with the {\it Poisson brackets}
\begin{gather*}
\boldsymbol{\{}F, \, G\boldsymbol{\}}:= \frac{\partial F}{\partial Q^{\text{\sffamily{A}}}}\frac{\partial G}{\partial P_{\text{\sffamily{A}}}} - \frac{\partial F}{\partial
P_{\text{\sffamily{A}}}}\frac{\partial G}{\partial Q^{\text{\sffamily{A}}}}.
\end{gather*}
From a~more geometrical perspective, Poisson brackets are well-known to be recastable in terms of a~symplectic
form~\cite{Arnol'd}.
These notions readily extend to f\/ield theories by upgrading to suitable functionals and including suitable integrals
over one's notion of space.

Moreover, passage from a~Lagrangian perspective to a~Hamiltonian one can be nontrivial.
Furthermore, it is the Hamiltonian perspective which possesses a~systematic treatment of constraints, due to
Dirac~\cite{Dirac, HT92}.
The Hamiltonian perspective additionally of\/fers a~more direct link to quantum theory.
The above nontriviality is due to the array
\begin{gather*}
{\partial^2 \text{\sffamily L}}/ \partial \dot{Q}^{\text{\sffamily{A}}}\partial\dot{Q}^{\text{\sffamily{A}}^{\prime}} \qquad \left(= \partial P_{\text{\sffamily{A}}^{\prime}}/{\partial \dot{Q}^{\text{\sffamily{A}}}} \right)
\end{gather*}
-- associated with the Legendre transformation~-- in general being non-invertible, by which the momenta cannot be
independent functions of the velocities.
Thus there are relations ${\cal C}_{\text{\sffamily{C}}}(Q^{\text{\sffamily{A}}},
P_{\text{\sffamily{A}}}) = 0$ between the momenta; these are standardly termed {\it constraints}.
(In this review, constraints are highlighted by exclusive use of the calligraphic font.)   Moreover, the above array
also features in the reformulation of the Euler--Lagrange equations as
\begin{gather*}
\ddot{Q}^{\text{\sffamily{A}}^{\prime}}   {\partial^2 \text{\sffamily L}}/{\partial
\dot{Q}^{\text{\sffamily{A}}}\partial\dot{Q}^{\text{\sffamily{A}}^{\prime}} = {\partial
\text{\sffamily L}}/ \partial Q^{\text{\sffamily{A}}}} - \dot{Q}^{\text{\sffamily{A}}^{\prime}}
  {\partial^2 \text{\sffamily L}}/ \partial {Q}^{\text{\sffamily{A}}}\partial\dot{Q}^{\text{\sffamily{A}}^{\prime}}.
\end{gather*}
Due to this, the noninvertibility has additional signif\/icance as accelerations not being uniquely determined by~$Q^A$,
$\dot{Q}^{\text{\sffamily{A}}}$.\footnote{For simplicity, this review's   range of physical theories
restricts itself to no higher than second-order theories.}

Constraints are usefully classif\/ied in a~number of ways, including the following due to Bergmann and
Dirac~\cite{Dirac, HT92}.

{\it Primary constraints} arise purely from the form of the Lagrangian; these are the relations between the
momenta by which the above-mentioned Legendre transformation maps onto only a~submanifold of the full phase space.

{\it Secondary constraints}, on the other hand, arise via use of the equations of motion.
One intuitively valuable case of this concerns constraints arising from the propagation of existing constraints using
the equations of motion.

{\it Weak equality} is equality up to additive functionals of the constraints; this holds on the {\it
constraint surface} (def\/ined as the surface within phase space where the totality of the constraints vanishes).

{\it First-class} constraints are then those whose classical brackets with all the other constraints vanish
weakly; these are indexed by~$\text{\sffamily F}$.
This can also be described in terms of no new entities~-- constraints or further kinds mentioned below~-- arising from
the bracket operation acting on~${\cal C}_{\text{\sffamily{F}}}$ and a~general ${\cal
C}_{\text{\sffamily{C}}}$.
Geometrically, these are characterized as the brackets that vanish on the version of constraint surface upon which all
f\/irst-class constraints vanish.
Ab initio, the classical brackets involved are Poisson brackets.

{\it Second-class} constraints are then simply def\/ined by exclusion as those constraints that fail to be
f\/irst-class.

Moreover, one can always in principle handle second-class constraints by passing from Poisson to {\it Dirac brackets},
\begin{gather*}
\boldsymbol{\{}F,  G\boldsymbol{\}}{}^{\boldsymbol{*}}:= \boldsymbol{\{}F,   G\boldsymbol{\}} -
\boldsymbol{\{}F,  {\cal C}_{\text{\sffamily{S}}}\boldsymbol{\}} \boldsymbol{\{}{\cal
C}_{\text{\sffamily{S}}},   {\cal C}_{\text{\sffamily{S}}^{\prime}}\boldsymbol{\}}^{-1}
\boldsymbol{\{}{\cal C}_{\text{\sffamily{S}}^{\prime}},  G\boldsymbol{\}}.
\end{gather*}
Here the~--1 denotes the inverse of the given matrix whose $\text{\sffamily S}$ indices index irreducibly~\cite{Dirac,HT92} second-class constraints.
({\it Irreducibly} here refers to these constraints not being combineable in any manner so as to separate out any
further functionally-independent f\/irst-class constraints.)   Then the classical brackets role played ab initio by the
Poisson brackets gets taken over by the Dirac brackets.
Moreover, e.g.~\cite{HT92, Sni} exposit how Dirac brackets can be viewed geometrically as a~more reduced formulation's
version of Poisson brackets.
The particular Dirac brackets formed once no second-class constraints remain illustrates the concept of `f\/inal classical
brackets' forming a~`f\/inal classical brackets algebra' of constraints.
(This is in contrast with na\"{\i}ve Poisson brackets as an `incipient' notion of bracket.)
First-class constraints use up 2~degrees of freedom each; second-class, only~1.

Some constraints are regarded as gauge constraints; however exactly which constraints these comprise is disputed.
What is agreed upon is that second-class constraints are not gauge constraints; all gauge constraints use up two degrees of freedom.
Dirac~\cite{Dirac} conjectured a~forteriori that all f\/irst-class constraints are gauge constraints\footnote{This is in
Dirac's sense of `gauge constraint' as per Section~\ref{1.3}.},   so using up 2 degrees of freedom would then
conversely imply being a~gauge constraint.
However Section~\ref{K-B} outlines how this conjecture has been refuted, alongside various other perspectives on the status
of gauge constraints.

{\it Gauge-fixing conditions} $\text{F}_{\text{\sffamily{X}}}$ may then be applied to whatever
gauge theory (though one requires the f\/inal answers to physical questions to be gauge-invariant).
These are a~means of removing gauge freedom by f\/ixing a~choice of gauge, though physical answers are required to end up
in gauge-invariant form.

As a~f\/inal remark, second-class constraints can always in principle\footnote{To~\cite{HT92}'s precursor
statement, I add the caveat `locally',   because gauge-f\/ixing conditions themselves in general are not global entities.}
be handled by alternatively thinking of them as `already-applied' gauge f\/ixing conditions that can be recast as
f\/irst-class constraints by adding suitable auxiliary variables to one's conf\/iguration space or phase space.
By doing this, a~system with f\/irst- and second-class constraints can be turned into a~more redundant description of
a~system with just f\/irst-class constraints.
Sets of f\/irst-class constraints obtained in this way are known as {\it effective constraints}~\cite{BT91}.

\subsection{Examples of constraints in theoretical physics}\label{1.2}

Most of the theories given here are used as recurring examples in this review; using multiple examples in reviews is in
the tradition of Isham~\cite{I93} and Kucha\v{r}~\cite{Kuchar92}.
I enumerate the example theories and models in this review with f\/ixed example numbers~0 to~10 to keep these recurrences
manifest.

{\bf Example 1.} Electromagnetism in vacuo has the\footnote{Some notation for this subsection is as follows.
I use capital Latin indices for particle labels or internal indices, depending on context.
$A_i$ is the electromagnetic vector potential with conjugate momentum $\pi^i$.
$\underline{q}^I$ are particle positions with conjugate momenta $\underline{p}_I$ and masses~$m_I$.
The 4-$d$ spacetime is the pair $(\mathfrak{m}, g_{\mu\nu})$.
Here $\mathfrak{m}$ is the spacetime topological manifold and $g_{\mu\nu}$ is a~metric that
provides this with semi-Riemannian geometrical structure.
$g_{\mu\nu} = g_{\mu\nu}(X^{\rho})$, for $X^{\rho}$ spacetime coordinates.
The 3-$d$ spaces are pairs $(\Sigma, h_{ij})$ for a~f\/ixed topological manifold~$\Sigma$.
Thus such dynamical study restricts $\mathfrak{m}$
to be of the simple form $\Sigma \times I$ for~$I$ some kind of interval in $\mathbb{R}$.
Moreover, this f\/ixed spatial topological space is taken in this review to be a~compact without boundary one.
Finally~$\Sigma$ additionally comes equipped with suitable dif\/ferential and metric structure.
$h_{ij} = h_{ij}(x^k)$, for $x^k$ spatial coordinates, is a~spatial metric, with determinant~$h$, covariant derivative
$D_{i}$, Ricci scalar $R = R(x^e; h_{fg}]$, and conjugate momenta~$p^{ij}$ with trace~$p$.}
\begin{gather*}
(\text{Gauss constraint})
\qquad
{\cal G}:= \partial_i \pi^i = 0.
\end{gather*}
This arises from variation with respect to the electromagnetic potential~$\Phi$.
One also has $\pi_{\Phi} = 0$, for $\pi_{\Phi}$ the momentum conjugate to~$\Phi$.
These are both f\/irst-class, and use up 2 degrees of freedom each, so one passes from $A_i$,~$\Phi$ and their conjugate
momenta's redundant $4 \times 2$ phase space degrees of freedom per space point to just~$2 \times 2$.
This is in accord with electromagnetic waves consisting of just~2 transverse modes.
This ${\cal G}$ is uncontroversially a~gauge constraint, associated with the ${\rm U}(1)$ group.
Further Gauss constraints that share these conceptual properties feature in many other theories.
Examples of such are in 1) pure Yang--Mills theory (with internal index~$I$: ${\cal G}_I$),
2)~the scalar and fermionic gauge theories that one can associate with each of electromagnetism and Yang--Mills theory.
(See~\cite{Weinberg+} for more details of~1) and~2)).

{\bf Example 2.} Barbour--Bertotti's~\cite{FileR, BB82} scaled {\it relational particle mechanics} has the
\begin{gather*}
(\text{zero total momentum constraint}) \qquad \underline{\cal P}:=
\sum_{I=1}^{N} \underline{p}_I = 0,
\qquad
\text{and}
\\
(\text{zero total angular momentum constraint}) \qquad \underline{\cal L}:=
\sum_{I=1}^{N} \underline{q}^I \times
\underline{p}_I = 0.
\end{gather*}
Here~$I$ runs over particle labels 1 to~$N$.
These constraints from variation with respect to some translational and rotational auxiliary variables respectively;
relatedly, these constraints generate the Euclidean group of translations and rotations.
They are f\/irst-class and use up 2 degrees of freedom per constraint degree of freedom.
Thus one passes from a~redundant conf\/iguration space $\mathbb{R}^{Nd}$ (in dimension~$d$) to a~reduced conf\/iguration
space $\mathbb{R}^{Nd}/\text{Eucl}(d)$.
This amounts to removing Newton's absolute space from mechanics.
Note that these are again not internal gauge constraints, but they are uncontroversially gauge constraints once more.
Molecular physics has similar classical kinematics in its zero angular momentum case; the nonzero angular momentum case,
however, has a~more complicated f\/ibre bundle structure (consult~\cite{LR97} if interested in this dif\/ference).

\begin{figure}[ht]
\centering

\includegraphics{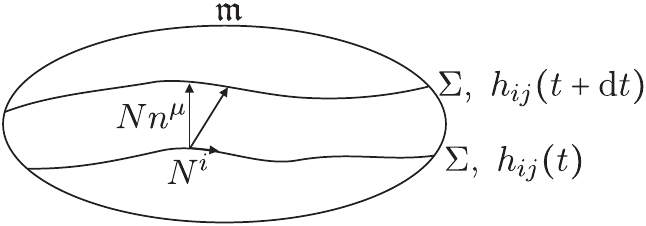}

\caption{ADM split of spacetime $\mathfrak{m}$ with respect to
spatial hypersurfaces~$\Sigma$.
$n^{\mu}$ is the normal to the hypersurface,~$N$ is the {\it lapse} (time elapsed) and $N^i$ is the {\it shift} (point
identif\/ication map).
Together, these form a~strutting: how to f\/it together adjacent hypersurfaces within spacetime.
For later use, 1)~the normal to the spatial hypersurface~$\Sigma$ then takes the computational form $n^{\mu} = [N^{-1},
- N^{-1}N^i]$.
2)~$N^{\mu}:= [N, N^i]$ is the spacetime 4-vector of auxiliaries, with conjugate momenta $P_{\mu}$.}
\label{Fig-1}
\end{figure}

{\bf Example 3.} Arnowitt--Deser--Misner (ADM)'s geometrodynamical formulation of GR \linebreak (Fig.~\ref{Fig-1}) involves the
\begin{gather*}
(\text{momentum constraint}) \qquad {\cal M}_i:= -2D_jp^j{}_i = 0.
\end{gather*}
This arises from variation with respect to ADM's shift $N^i$.
${\cal M}_i$ is f\/irst-class, and uses up 2 degrees of freedom per space point.
It is also uncontroversially a~gauge constraint, with the {\it spatial diffeomorphisms} Dif\/f($\Sigma$) as the
corresponding gauge group.

The feature of using up degrees of freedom in pairs also applies to the GR
\begin{gather*}
(\text{Hamiltonian constraint}) \qquad {\cal H}:= \big\{p_{ij}p^{ij} - {p^2}/{2}\big\}\big/\sqrt{h} - \sqrt{h}R = 0.
\end{gather*}
In the ADM formulation of GR this also arises as a~secondary constraint from variation with respect to the lapse~$N$.
On the other hand, in the Baierlein--Sharp--Wheeler or related forms of GR~\cite{BSW, BFOA}, ${\cal H}$ arises rather as
a~primary constraint corresponding to the action being reparametrization invariant along the lines of~\eqref{S-RI}.
The same is true for relational particle mechanics'
\begin{gather*}
(\text{energy constraint}) \qquad {\cal E}:= \sum_{I=1}^{N}
p_I^2/2m_I + V\big(\underline{q}^I\big) = E,
\end{gather*}
as argued below
($V$ is here the potential function).
Thus the below two examples illustrate that the primary-secondary distinction is artif\/icial in that it is malleable~by
change of formalism.
Thus we will avoid that distinction in this review other than possibly pointing out others' claims that concern it (and
counterexamples).

Example~2 relational particle mechanics' energy constraint ${\cal E}$ arises as a~primary constraint from
their Jacobi-type action~\cite{FileR, BB82, Lanczos}\footnote{Here   ${\underline{a}}$ and ${\underline{b}}$ are
translational and rotational auxiliaries respectively.
$m_I$ are particle masses,~$E$ is the total energy and~$T$ is the kinetic term.}
\begin{gather}
\text{\sffamily S} = \int \text{\sffamily L}   \textrm{d}\lambda = 2\int\sqrt{T\{E - V\}}\textrm{d}\lambda,
\nonumber
\\
T = \delta_{IJ}m_I\big\{\dot{\underline{q}}^I - \dot{\underline{a}} - \big\{\dot{\underline{b}}
 \times  \underline{q}^I\big\}\big\} \big\{\dot{\underline{q}}^J - \dot{\underline{a}} -
\big\{\dot{\underline{b}}  \times  \underline{q}^J\big\}\big\}/2.
\label{S-RI}
\end{gather}
Moreover in this case the way the purely quadratic form of the Lagrangian causes the constraint to arise is in close
analogy to Pythagoras' theorem/direction cosines summing to one.

{\bf Example 4.} Then the Baierlein--Sharp--Wheeler or related formulations~\cite{BSW, BFOA} of GR have the GR
Hamiltonian constraint ${\cal H}$ arise as one primary constraint per space point in close analogy to Example~4  working.

\subsection{Interlude: notions of gauge theory}\label{1.3}

One conceptually useful way of introducing gauge theory\footnote{We take this to have a~wider meaning than just the
typical gauge theories of particle physics.
It covers also e.g.\ the gauge theories in molecular physics~\cite{LR97}, relational particle mechanics~\cite{FileR, BB82}, cosmological
perturbation theory~\cite{Bardeen, BM}, and those associated with various kinds of dif\/feomorphisms (see
e.g.~\cite{ Diffs2, Diffs1}).}   is by letting $\mathfrak{g}$ be a~group of transformations held to be
physically redundant that acts on $\mathfrak{q}$ (or sometimes, when specif\/ied in this review, on phase
space).
This group (`gauge group'), the constraints and gauge theory can then be inter-related as follows.

1) The mathematically-disjoint auxiliaries $g^{\text{\sffamily{G}}}$ are $\mathfrak{g}$-auxiliaries that encode the group action of~$\mathfrak{g}$.
(Mathematically-disjoint means like~$\Phi$ not being part of a~larger tensorial package in the sense that the
longitudinal piece of $A^a$ is part of the larger entity $A^a$ itself.)   Then at least in the set of examples given
above, f\/irst-class secondary constraints arise from variation with respect to mathematically-disjoint auxiliary
variables.
Furthermore, the ef\/fect of this variation is to additionally use up part of an accompanying mathematically coherent
block that however only contains partially physical information; this is clear in the above discussions of
electromagnetism.

2) The disjoint auxiliary variables are moreover often in correspondence with a~group of redundancies
$\mathfrak{g}$.
Variation with respect to the mathematically disjoint auxiliary variable $g^{\text{\sffamily{G}}}$ produces
the gauge constraints, denoted ${\cal GAUGE}_{\text{\sffamily{G}}}$ for clarity.
We then do not need to isolate the latter for many purposes: E.g.\
in the case of electromagnetism, the Gauss constraint ${\cal G}$ associated with this pair already arises from varying
with respect to the auxiliary~$\Phi$.
It is much more convenient to obtain ${\cal G}$ in this way because~$\Phi$ is a~mathematically isolated object and so
one can entirely straightforwardly vary with respect to it.
This is one reason why the most habitual~-- and sometimes the only possible~-- redundant formulations of gauge theories
are useful to work with.   Some theoreticians (contrast with Section~\ref{K-B-Motiv}) preclude from 2) the
reparametrization and refoliation groups on grounds that they are dynamically distinct.
The included groups, the constraints corresponding to which are f\/irst-class and linear~-- denoted ${\cal
LIN}_{\text{\sffamily{L}}}$~-- fall under Barbour's best matching paradigm (outlined in Appendix~\ref{PoT}) and
correspond to redundancies in the instantaneous conf\/igurations\footnote{Thus the corresponding notion of gauge   is
instantaneous, i.e.~along the lines of Dirac's notion and not Bergmann's, as discussed below.}.

The precluded constraints are quadratic constraints (corresponding to strutting `orthogonal' to the instantaneous
conf\/igurations: see Fig.~\ref{Fig-1}).
Ways in which refoliation is more subtle than reparametrization are outlined in Section~\ref{1.7}.
In the commonly-encountered physical theories, the diagnostic for the precluded constraints is that they have quadratic
and not linear dependence in the constraints; hence I denote these by ${\cal QUAD}$.

Note that naming a~$\mathfrak{g}$ as a~candidate gauge group can be a~formalism-dependent rather than
theory-dependent statement.
On the one hand, it is at least in principle possible to rewrite a~theory possessing a~gauge freedom in terms of true
dynamical degrees of freedom alone (see also Section~\ref{D-B-Motiv}).
(This `in principle' does not imply that the equations one would need to solve to do so can be solved.)   On the other
hand, variable numbers of auxiliary degrees of freedom can be added, and some such cause dif\/ferent gauge constraints to
appear or to recast existing non-gauge constraints as gauge ones.
These considerations parallel those for removing second-class constraints in Section~\ref{1.1}.
Moreover notions~1) and~2) can be applied, for a~subset of theories, to the subset of constraints that are linear.
I.e.\
that subset of theories for which ${\cal QUAD}$ (or any other nonlinear constraint) is not an integrability of the
linear constraints ${\cal LIN}_{\text{\sffamily{L}}}$ (see Section~\ref{1.7} for a~counterexample).
Thus here ${\cal LIN}_{\text{\sffamily{L}}}$ constitutes a~subalgebraic structure (taken to cover both
subalgebras and subalgebroids, see Section~\ref{1.7}) of constraints.

There are of course further conceptualizations of gauge and of gauge theory~\cite{Bleecker, Rovelli13, Weinberg,Weinberg+}.
For instance, one can do so in terms of the presence of free functions in the solution of the equations of motion or of
making global symmetries local.

A further, older distinction between dif\/ferent uses of the word `gauge' concerns what a~theory is a~gauge theory of,
e.g., $Q^{\text{\sffamily{A}}}$ alone, $Q^{\text{\sffamily{A}}}$ and
$P_{\text{\sffamily{A}}}$, whole paths, or histories.
Bergmann's early position~\cite{Bergmann61} was that that gauge theory concerns {\it whole paths} (dynamical
trajectories): {\it path-gauge} notion.
This is in contrast to Dirac's perspective~\cite{DiracObs, Dirac} that gauge theory concerns {\it data at a~given time}:
{\it data-gauge} notion (called `D' by Bergmann and Komar~\cite{BK72}, albeit that stood for `Dirac' rather than for
`data')\footnote{N.B.\
path-gauge and data-gauge are   def\/initions of {\it notions of} gauge, as opposed to particular {\it choices of} gauge
within a~particular notion of gauge such as Coulomb gauge or Lorenz gauge for electromagnetism).
In this review, I also take `history' to mean more than just a~dynamical path; at the quantum level these are paths that
are furthermore decorated with projection operators.
Also N.B.\
that this review's namings are preferentially based on each entity's conceptual content~-- {\it true name}~-- rather
than on e.g.\ the name of who discovered it, or on how the entity was once thought about prior to developments in its conceptual
understanding.
Of course, upon f\/irst introduction of such terms, I give what aliases they are or have been known by.}.
I.e.\ gauge group action in spacetime versus spatial/conf\/igurational/canonical settings.

\subsection{More examples of constraints toward quantum gravitational theories}\label{1.4}

{\bf Example 5.} Ashtekar Variables formulation of GR's constraints are\footnote{I present just the complex case for
simplicity.
$E^i_I$ is the ${\rm SU}(2)$ equivalent of electric f\/ield f\/lux.
$F_{ijI}$ is the corresponding f\/ield strength.
$E^i_I$ is now also geometrically a~particular kind of {\it bein} (`square root' of the spatial metric conf\/iguration
variable), and yet has been recast as a~momentum variable by canonical transformation.}
\begin{gather*}
{\cal G}_I:= D_{i} E^{i}_I = 0,
\qquad
{\cal M}_i:= E^{jI} F_{jiI} = 0,
\qquad
{\cal H}:= \epsilon_{IJK} E^{iI}E^{jJ}F_{ij}^K = 0.
\end{gather*}

{\bf Example 6.} Proca theory (the massive counterpart of electromagnetism) is a~simple example of a~theory with
a~second-class constraint~\cite{P78}
\begin{gather*}
{\cal C}:= \partial_i \pi^i + m^2\Phi = 0.
\end{gather*}
This indeed uses up only~1 degree of freedom, so this theory has 1 more physical mode than electromagnetism itself.
Gravitational theories with second-class constraints include 1)~Einstein--Cartan theory~\cite{K78},
2)~Einstein--Dirac theory (i.e.~GR with spin-1/2 fermion matter as required for a~full model of the known f\/ields of
nature)~\cite{DEath},
3)~supergravity~\cite{DEath, Des, FV, P78, T77} (on the one hand supersymmetric particles are being sought for at the
LHC, and on the other hand this review reveals a~number of further ways in which supergravity dif\/fers from GR).

{\bf Example 7.} Locally-Lorentz constraints in f\/irst-order formulations have constraints ${\cal J}_{AB}$ (and
conjugate; the capital indices here are specif\/ically 2-spinor indices).
These occur in e.g.\ in Einstein--Dirac theory and supergravity.

{\bf Example 8.} Supersymmetric constraints, a~particular case of which in gravitational theories are supergravity's
constraint~${\cal S}_A$ (and conjugate).

See e.g.~\cite{DEath} for explicit forms for ${\cal J}_{AB}$ and ${\cal S}_A$ (these details are not required
for this review, which only makes use of the form of the algebraic structure of the constraints).

\subsection{Dif\/ferent kinds of notions of equality in the principles of dynamics}\label{1.5}

We already explained weak equality in Section~\ref{1.1}: `up to functionals of the f\/irst-class constraints'.
`Strong equality' means equality in the usual sense\footnote{E.g.~\cite{HT92} give a~form for this as a~linear
combination of constraints.
I however retain the general def\/inition for its subsequent use in studying global ef\/fects.}.     E.g.\
Isham~\cite{I93} points out the possibility of making the strong equality demand in def\/ining notions of beables.
Finally, Batalin and Tyutin~\cite{BT91} consider equality up to ef\/fective constraints, denoted by the symbol
$\triplesim$.
I term this {\it effective weak equality}.
Moreover, one person's ef\/fective formulation could have been written down ab initio by another as a~formulation
happening to have no second-class constraints, so this is not so large a~distinction.

\subsection{Dif\/ferent kinds of brackets}\label{1.6}

We have already encountered the Poisson brackets and the Dirac brackets in Section~\ref{1.1}.
A~distinct generalization of the Poisson bracket~-- to mixtures of bosonic and fermionic species~-- is Poisson bracket
here generalizes to the {\it Casalbuoni brackets}~\cite{Casalbuoni}
\begin{gather*}
\boldsymbol{\{}F,   G\boldsymbol{\}}_{\text{{\bf C}}}:= \frac{\partial F}{\partial
Q^{\text{\sffamily{A}}}}\frac{\partial G}{\partial P_{\text{\sffamily{A}}}} -
(-)^{\epsilon_F\epsilon_G} \frac{\partial G}{\partial Q^{\text{\sffamily{A}}}}\frac{\partial F}{\partial
P_{\text{\sffamily{A}}}}.
\end{gather*}
Here $\epsilon_{\text{\sffamily{A}}}$ is the Grassmann parity of species $\text{\sffamily A}$: 1 for bosons
and~$-1$ for fermions.
This also readily generalizes to f\/ield-theoretic form.
It obeys the {\it Grassmannian generalization of the Jacobi identity},
\begin{gather*}
\boldsymbol{\{}\boldsymbol{\{}F ,   G \boldsymbol{\}}_{\text{{\bf C}}} ,   H \boldsymbol{ \}}_{\text{{\bf C}}}(-1)^{\epsilon_F\epsilon_H} + \text{cycles} = 0.
%\label{Grass-Jac}
\end{gather*}

The quantum commutator counterpart of the above types of brackets is covered in Section~\ref{1.9}.
See Section~\ref{Var-Bra} for yet further types of classical and quantum brackets.

\subsection{Algebraic structures resulting from the introduction of brackets}\label{1.7}

Given a~type of bracket, there is the additional issue of mathematical type of the algebraic structure formed~by
entering the theory's constraints into that type of brackets.
It is well known that if the right hand side is of the form of a~sum of (structure constants) $\times$ constraints, the
brackets of constraints constitute a~{\it Lie algebra}.
However, if instead structure functions materialize, one has a~more general structure termed an {\it
algebroid}~\cite{BojoBook, Algebroid1, Algebroid2}.
This clearly structurally precedes the notions of beables and of beables algebraic structures, through using a~strict
subset of the structures that this requires.

{\bf Example 1.} Gauss constraints form Lie algebras.
These can be Abelian (electromagne\-tism)\footnote{This is given with smearing functions   $\chi(z)$ and $\zeta(z)$.
More generally, $({\cal C}_{\text{\sffamily{Z}}}| A^{\text{\sffamily{Z}}}):= \int d^3z \, {\cal
C}_{\text{\sffamily{Z}}}(z^i; h_{jk}]   A^{\text{\sffamily{Z}}}(z^i)$ denotes an `inner product' notation
for the smearing of a~$\text{\sffamily{Z}}$-tensor density valued constraint ${\cal
C}_{\text{\sffamily{Z}}}$ by an opposite-rank $\text{\sffamily{Z}}$-tensor smearing with no density
weighting, $A^{\text{\sffamily{Z}}}$.}
\begin{gather*}
\boldsymbol{\{}({\cal G}|\chi),   ({\cal G}|\zeta)\boldsymbol{\}} = 0,
\end{gather*}
or non-Abelian (Yang--Mills theory)
\begin{gather*}
\boldsymbol{\{}({\cal G}_I|\chi^I),   ({\cal G}_J|\zeta^J)\boldsymbol{\}} = f_{IJ}{}^K ({\cal
G}_K|\chi^I\zeta^J)
\end{gather*}
for Lie algebra structure constants $f_{IJ}{}^K$.

{\bf Example 2.} 3-$d$ relational particle mechanics has
\begin{gather*}
\boldsymbol{\{}{\cal P}_i,   {\cal P}_j\boldsymbol{\}} = 0,
\qquad
\boldsymbol{\{}{\cal
L}_i,   {\cal L}_j\boldsymbol{\}} = \epsilon_{ij}{}^k{\cal L}_k,
\qquad
\boldsymbol{\{}{\cal P}_i,   {\cal L}_j\boldsymbol{\}} = \epsilon_{ij}{}^k{\cal P}_k.
\end{gather*}
The meanings of these are, respectively, that the ${\cal L}_i$ form an ${\rm SO}(3)$ subalgebra of rotations, the ${\cal P}_i$
form an Abelian subalgebra of translations, and ${\cal P}_i$ is a~good ${\cal L}_i$ vector.
All other Poisson brackets for this model are zero.

{\bf Example 3  or 4.} Spatial dif\/feomorphisms by themselves form an inf\/inite-dimensional Lie algebra,
\begin{gather}
\boldsymbol{\{} (\mathcal M_i | L^i ) , (\mathcal M_j | M^j  ) \boldsymbol{\}} = (\mathcal M_k
|\boldsymbol{[} L , M \boldsymbol{]}|^k) .
\label{MM}
\end{gather}
Here $L_i(z)$ and $M_i(z)$ are vectorial smearing functions and $|\boldsymbol{[} \, , \, \boldsymbol{]}|$ is the
standard Lie bracket.

{\bf Example 3  or 4.} In the case of full GR as geometrodynamics, the constraints' Poisson brackets
are~\cite{Dirac}~\eqref{MM},
\begin{gather}
\boldsymbol{\{} (\mathcal H | K  ) ,   (\mathcal M_i | L^i  ) \boldsymbol{\}} = (\pounds_{L} \mathcal H | K  ),
\label{HM}
\\
\boldsymbol{\{} (\mathcal H | J  ) ,   (\mathcal H | K  )\boldsymbol{\}} = (\mathcal M_i h^{ij} | {J}
\overleftrightarrow{\partial}_i {K})  .
\label{HH}
\end{gather}
Here $J(z)$, $K(z)$ are scalar smearing functions, $\pounds$ is the well-known Lie derivative of dif\/ferential
geometry~\cite{Stewart}, and $X \overleftrightarrow{\partial}_i Y:= (\partial_i Y) X - Y \partial_i X$ (a notation
familiar from QFT).

\eqref{HM}~means that ${\cal H}$ is a~`good object' (a scalar density) under spatial dif\/feomorphisms Dif\/f($\Sigma$).
\eqref{HH}, however, is more complicated in both form and meaning.
Firstly, it uncontroversially means that the linear ${\cal M}_i$ arises as an integrability of the quadratic ${\cal H}$.
Secondly, the presence of $h^{ij}(h_{kl})$ on its right-hand side is a~structure function.
Thus this bracket renders the overall algebraic structure more complicated than a~Lie algebra.
It is, rather, an algebroid: specif\/ically the {\it Dirac algebroid}~\cite{Dirac, BojoBook}.
This is entirely unlike the {\it unsplit} GR's spacetime dif\/feomorphisms Dif\/f($\mathfrak{m}$) which form
a~genuine Lie algebra paralleling that of Dif\/f($\Sigma$).

${\cal H}$'s distinction from GR theories' linear constraints has further fuel than ${\cal E}$'s from
relational particle mechanics linear constraints, as follows.
i) Refoliation invariance is a~{\it hidden invariance}.
(This is as opposed to an invariance that is manifest in the canonical formulation itself.
In particular, one needs to foliate spacetime in order to have a~canonical formulation, and one cannot directly see
refoliation invariance within any particular foliation.)   ii)~\eqref{HH} implies that GR's linear momentum
constraint ${\cal M}_i$ is an integrability condition that follows from the existence of ${\cal H}$.
Thus it ceases to be possible to consider ${\cal QUAD}$ and ${\cal LIN}_{\text{\sffamily{L}}} $ piecemeal in
generally-relativistic theories.
iii)~It is specif\/ically the presence of ${\cal H}$ that causes the algebraic structure of these constraints to be an
algebroid.
Structure functions are needed to accommodate the variety of possible foliations; see e.g.~\cite{IK85I, IK85II, LeeWald, T73} for further discussions of the `group action' involved.

{\bf Example 5.} The Ashtekar variables' algebraic structure of constraints~\cite{A91, Thiemannbook} is much like
geometrodynamics', but with an extra Gauss-type constraint included.
This further includes a~bracket between two ${\rm SU}(2)$ Yang--Mills--Gauss constraints (these simply commute with the two
other constraints).
It is the bracket of two ${\cal H}$'s that continues to cause dif\/f\/iculty, and for reasons unchanged from the
geometrodynamical case's.

{\bf Examples 6 and 7.} See e.g.~\cite{DEath, VM10} for what is known about the Einstein--Dirac and supergravity
constraint algebras.
In particular, supergravity is an example of a~subset of the linear constraints~-- the supersymmetry constraints~--
having the supergravity counterpart of the quadratic ${\cal H}$ as {\it their} integrability condition.
This follows from~\cite{DEath, T77}
\begin{gather}
\boldsymbol{\{} ({\cal S}_{A}|\theta^{A}) ,   (\overline{\cal S}_{A^{\prime}} |
\overline\theta{}^{A^{\prime}}) \boldsymbol{\}}{}^{\boldsymbol{*}}_{\text{{\bf C}}}
\propto
i (\gamma^{AA^{\prime}}|{\cal H}_{AA^{\prime}}) + \text{terms with ${\cal J}_{AB}$ or its conjugate as a~factor}.
\label{SSH} \hspace{-5mm}
\end{gather}
Here ${\cal H}_{AA^{\prime}}:= n_{AA^{\prime}}{\cal H}_{\perp} + e_{AA^{\prime}}{}^i{\cal M}_i$ packages together
the supergravity Hamiltonian and momentum constraints using the normal~$n$ and spinor-valued 1-form~$e$.
(Less importantly, $\theta^A$ and $\overline{\theta}{}^{A^{\prime}}$ are fermionic smearing functions, whereas
$\gamma^{AA^{\prime}}(\theta^B,  \overline{\theta}{}^{B^{\prime}})$ is some composite of these that is itself
another smearing function.)   \eqref{SSH} is the basis for a~number of signif\/icant new results in this review.

\eqref{SSH} can furthermore be interpreted~\cite{T77} in terms of ${\cal S}_A$ being a~square root of ${\cal
H}$ in parallel to how the Dirac operator is well known to be a~square root of the Klein--Gordon one.
This may provide reasons why ${\cal H}$ is, after all, not so fundamental.
However, it should be cautioned that whereas Dirac's corresponding fermions were observationally vindicated, this is not
the case to date as regards superpartner particles.
This can be taken as a~limitation on arguing against the fundamentality of quadratic constraints like ${\cal H}$ on the
grounds of their being supplanted in supersymmetric theories.

To sum up, schematically, for a~theory with constraints, the constraint algebra is
\begin{gather}
|\boldsymbol{[} {\cal C}_{\text{\sffamily{F}}},   {\cal C}_{\text{\sffamily{F}}^{\prime}} \boldsymbol{]}| \approx 0.
\label{C-C}
\end{gather}
Indexing these constraints by $\text{\sffamily F}$'s indicates that they are f\/irst-class.
Any second-class ones there were have been removed by one of the following.
a) Extension, in which case it indeed involves a~Poisson bracket, ef\/fective $\text{\sffamily E}$-index and ef\/fective
weak equality symbol $\triplesim$.
b) Passing to the Dirac bracket whilst leaving the $\text{\sffamily F}$-index and weak equality symbol untouched.

\subsection{Classical notions of beables}\label{1.8}

Finally, a~theory with constraints and brackets possesses a~further class of conceptually important objects: those that
form (usually) weakly zero brackets with the constraints,
\begin{gather}
|\boldsymbol{[} {\cal C}_{\text{\sffamily{F}}} ,  B_{\text{\sffamily{B}}} \boldsymbol{]}|
\approx 0.
\label{B-C}
\end{gather}
Comparing~\eqref{C-C} and~\eqref{B-C} implies that the ${\cal C}_{\text{\sffamily{F}}}$ themselves are in
some sense beables.
However, already ${\cal C}_{\text{\sffamily{F}}} \approx 0$, so we are really looking for further quantities
that are not trivial in this way.
Let us call these other quantities {\it proper beables}; the rest of the article will always mean `proper beables'
whenever it says `beables'.
Together,~\eqref{C-C}, \eqref{B-C} and closure of beables carry no nontrivial algebraic structure at the level of weak
equality than just the closure of beables.
This is because they are just the conditions for a~direct product with the weakly-Abelian constraint algebra.

In all cases beables themselves are to close as an algebraic structure under the same type of bracket that they are
def\/ined by.
E.g.
if $B_{\text{\sffamily{B}}}$ are beables in the sense of~\eqref{basic-obs}, then $|\boldsymbol{[}B_{\text{\sffamily{B}}},   B_{\text{\sffamily{B}}^{\prime}}\boldsymbol{]}|$ are too.
I.e.\
this bracket object itself obeys property~\eqref{basic-obs}.
This is by two uses of~\eqref{basic-obs} in the Jacobi identity with two~$B$'s and one ${\cal C}$:
\begin{gather}
|\boldsymbol{[}{\cal C}_{\text{\sffamily{F}}},   |\boldsymbol{[}B_{\text{\sffamily{B}}}
,   B_{\text{\sffamily{B}}^{\prime}} \boldsymbol{]}|   \boldsymbol{]}| = -
|\boldsymbol{[}B_{\text{\sffamily{B}}},   |\boldsymbol{[}B_{\text{\sffamily{B}}^{\prime}}, {\cal C}_{\text{\sffamily{F}}} \boldsymbol{]}| \boldsymbol{]}| -
|\boldsymbol{[}B_{\text{\sffamily{B}}^{\prime}},   |\boldsymbol{[}{\cal C}_{\text{\sffamily{F}}}, B_{\text{\sffamily{B}}} \boldsymbol{]}| \boldsymbol{]}| = 0.
\label{BBC}
\end{gather}
On the other hand, two uses of~\eqref{basic-obs} in the less usual Jacobi identity with two ${\cal C}$'s and one~$B$,
\begin{gather}
|\boldsymbol{[}B_{\text{\sffamily{B}}},   |\boldsymbol{[}{\cal C}_{\text{\sffamily{F}}}
,   {\cal C}_{\text{\sffamily{C}}^{\prime}} \boldsymbol{]}| \boldsymbol{]}| = - |\boldsymbol{[}{\cal
C}_{\text{\sffamily{F}}},   |\boldsymbol{[}{\cal C}_{\text{\sffamily{F}}^{\prime}}, B_{\text{\sffamily{B}}} \boldsymbol{]}|   \boldsymbol{]}| - |\boldsymbol{ [}{\cal   C}_{\text{\sffamily{F}}^{\prime}},   |\boldsymbol{[}B_{\text{\sffamily{B}}},
{\cal C}_{\text{\sffamily{F}}} \boldsymbol{]}| \boldsymbol{]}| = 0,
\label{CCB}
\end{gather}
enforces the following.
A~particular notion of beables $B_{\text{\sffamily{B}}}$ corresponding to forming zero brackets with
a~subset of a~theory's constraints ${\cal C}_{\text{\sffamily{C}}}$ is only self-consistent if
$B_{\text{\sffamily{B}}}$ also forms zero brackets with $|\boldsymbol{[}{\cal C}_{\text{\sffamily{F}}} ,   {\cal C}_{\text{\sffamily{F}}^{\prime}}\boldsymbol{]}|$.    I.e.\
with the algebraic closure of that subset of constraints.
Thus one can only consistently adopt subsets of constraints that are furthermore subalgebraic structures with respect to
the given brackets.

Dif\/ferent notions of beables then concern dif\/ferent subalgebraic structures of constraints.

{\it Classical Dirac beables} (Section~\ref{D-B}) are functionals $D_{\text{\sffamily{D}}} =
\text{F}_{\text{\sffamily{D}}}[Q^{\text{\sffamily{A}}}, P_{\text{\sffamily{A}}}]$
that classical-bracket-commute with {\it all} of a~theory's f\/irst-class constraints ${\cal
C}_{\text{\sffamily{F}}}$ then using the Poisson, Dirac or the enlarged bracket for the purpose of assigning
beables
\begin{gather}
\boldsymbol{\{} {\cal C}_{\text{\sffamily{F}}},   D_{\text{\sffamily{D}}} \boldsymbol{\}}
\approx 0
\label{DirObs}
\end{gather}
is the standard form for this.
Batalin and Tyutin also introduced a~weak ef\/fective equality version~\cite{BT91}.
In the case in which time evolution is generated by a~constraint, Dirac beables are also known as {\it constants of the
motion} alias {\it perennials}~\cite{BF08,Ear01,Ear02, PG05, +Perennials1, +Perennials2,Kuchar93, K99, WuthrichTh}.
{\it True}~\cite{Rov91a, Rov91c, Rov91b} alias {\it complete observables/beables}~\cite{Rov02a, Rov02b} (which at
least~\cite{Thiemannbook} also terms evolving constant of the motion) are also a~notion of this kind.
They involve operations on a~system each of which produces a~number that can be predicted if the state of the system is
known.
See Section~\ref{D-B} for examples.

{\bf Note 1.} If there were second-class constraints, we would pass to Dirac or extended brackets, whence they are absent and
then def\/ine Dirac beables as before but in terms of this new bracket.

{\bf Note 2.} The name `constants of the motion' conventionally follows from a~generally-covariant (at least in
Henneaux and Teitelboim's sense~\cite{HT92}) theory's total Hamiltonian being of the form $\text{\sffamily H} =
\int_{\Sigma}\textrm{d} \Sigma \, \Lambda^{\text{\sffamily{F}}} {\cal C}_{\text{\sffamily{F}}}$
for multiplier coordinates $\Lambda^{\text{\sffamily{F}}}$.
Then
\begin{gather}
\textrm{d} D_{\text{\sffamily{D}}}/\textrm{d} t = \boldsymbol{\{} D_{\text{\sffamily{D}}},   \text{\sffamily H} \boldsymbol{\}}
=\boldsymbol{\{} D_{\text{\sffamily{D}}},
\int_{\Sigma}\textrm{d} \Sigma \, \Lambda_{\text{\sffamily{F}}}{\cal
C}^{\text{\sffamily{F}}}\boldsymbol{\}}
= \int_{\Sigma}\textrm{d} \Sigma \,
\Lambda_{\text{\sffamily{F}}}\boldsymbol{\{}D_{\text{\sffamily{D}}},  {\cal C}^{\text{\sffamily{F}}}\boldsymbol{\}} \approx 0,
\label{Fally}
\end{gather}
with the last equality following from~\eqref{DirObs}.
Thus it would appear that `nothing happens' (a type of frozen argument), though Section~\ref{Fall} attributes this to
a~fallacy.

{\bf Note 3.} 't Hooft~\cite{tHooft} used a~notion of `beables' that are conceptually disjoint from his notion of
`changeables'; as a~frozen notion, however, 't Hooft's notion of beables is more stringent than the notion of beables
used in this review.

On the other hand, classical {\it Kucha\v{r} beables}~\cite{BF08, Ear01, Ear02,Kouletsis, Kuchar93, K99, WuthrichTh}
(Section~\ref{K-B}) are functionals $K_{\text{\sffamily{K}}} = \text{F}_{\text{\sffamily{K}}}[Q^{\text{\sffamily{A}}},P_{\text{\sffamily{A}}}]$ that classical-brackets-commute with
all {\it linear} constraints
\begin{gather*}
\boldsymbol{\{} K_{\text{\sffamily{K}}} ,   {\cal LIN}_{\text{\sffamily{L}}}\boldsymbol{\}} \approx 0.
%\label{KObs}
\end{gather*}
Kucha\v{r} beables are more straightforward to construct; see Section~\ref{K-B} for examples.
These correspond to an uncontroversial if perhaps somewhat restrictive notion of gauge invariance.
Namely the one given in Section~\ref{1.3} in terms of the gauge group $\mathfrak{g}$ that corresponds
to the linear constraints ${\cal LIN}_{\text{\sffamily{L}}}$.
Kucha\v{r} beables are then gauge-invariant quantities in the `Dirac' sense familiar from electromagnetism and the
canonical formulation of particle physics.
Using Kucha\v{r} beables ref\/lects treating ${\cal QUAD}$ distinctly from ${\cal LIN}_{\text{\sffamily{L}}}$;
see Section~\ref{K-B} for further motivation for this.

N.B.\ that weak and ef\/fective-weak notions are tied to the uncontroversial notion of f\/irst-class constraints rather than to
gauge constraints.
In the case of standard canonical formulations GR, ef\/fective is equivalent to f\/irst-class, and so ef\/fective-weak is
equivalent to weak.

Classical {\it partial observables} are a~point of view that began with Rovelli's works~\cite{Carlip90, Carlip91b, Rov91a,Rov91c, Rov91b} though one might view~\cite{DeWitt62, PW83} as forerunners in some respects.
See also~\cite{Dittrich1, Dittrich2,Rov02a, Rovelli13} and the reviews~\cite{Rovellibook, Tambornino, Thiemannbook}.
Partial observables do not require commutation with any constraints.
Partial observables involve classical or QM operations on the system that produces a~number that is measurable but
possibly totally unpredictable even if the state is perfectly known (contrast with the def\/inition of total/Dirac
observables).
The physics then lies in considering pairs of these objects, with correlations between them encoding extractable purely
physical information.
I.e.\
correlations of two partial observables are predictable; in particular the value of a~partial observable~A subject to
another partial observable B taking a~particular value is predictable, in which case partial observable~B is playing
a~`clock' role.
It is not however clear exactly which partial observables correspond to realistic and accurate clocks.
Nor is it clear how a~number of other facets of the problem of time can be addressed via these~\cite{APOT2, FileR, I93, Kuchar92,
Kuchar93}.
What is clear is that the partial observables approach's correlations are themselves functions on the constraint surface
and commute with the constraints; as such they furnish complete or Dirac observables/beables, according to one's
interpretation.

Section~\ref{Fa-De} then covers local versus global notions of beables, and Section~\ref{Diff} covers Pons et al.
dif\/feomorphism-specif\/ic work~\cite{JP09,PS05,PSS97,PSS09b, PSS09, PSS10}.
The latter also covers how Bergmann observables/beables follow from his and various collaborators' position on the
notion of gauge~\cite{AB51, Bergmann61, BK72}.

\subsection{Quantum notions of beables}\label{1.9}

The quantum versions of the def\/initions of beables (see Section~\ref{Q-Beables} for more detail) involve
self-adjoint operators that form zero quantum commutators with the quantum constraints
\begin{gather}
\boldsymbol{[}\widehat{A} ,   \widehat{B} \boldsymbol{]} = \widehat{A}\widehat{B} - \widehat{B}\widehat{A},
\nonumber
\\
\text{quantum Dirac beables are: $\widehat{D}_{\text{\sffamily{D}}}$ such that}~\boldsymbol{[}\widehat{D}_{\text{\sffamily{D}}}, \widehat{\cal C}_{\text{\sffamily{F}}}\boldsymbol{]}\Psi = 0,
\label{QDB}
\\
\text{quantum Kucha\v{r} beables are: $\widehat{K}_{\text{\sffamily{K}}}$ such that}~\boldsymbol{[}\widehat{K}_{\text{\sffamily{K}}}, \widehat{\cal LIN}_{\text{\sffamily{L}}}\boldsymbol{]}\Psi = 0.
\label{QKB}
\end{gather}
Objects $\widehat{S}_{\text{\sffamily{M}}}$ obeying $\boldsymbol{[} \widehat{\cal QUAD} , \widehat{S}_{\text{\sffamily{M}}} \boldsymbol{]}\Psi = 0$ are conventionally termed {\it $S$-matrix quantities},
after the QM's scattering matrix for interaction processes.
Furthermore, these do not carry background-dependence connotations due to corresponding to `scattering processes' in
conf\/iguration space rather than in space itself.
Clearly then for quantum beables, Kucha\v{r} {\it and} $S$-matrix $\Rightarrow$ Dirac.
Quantum partial observables are def\/ined exactly as before too, though now `produce a~number' carries inherent
probabilistic connotations.

\subsection{The problem of beables}\label{1.10}

The problem of beables~\cite{APOT2, FileR, Dirac,I93, Kuchar92, Rovellibook} is but one facet of the problem
of time~\cite{APOT2, APOT, FileR, I93, Kuchar92} (see Appendix~\ref{PoT} for other facets of this).
It concerns that in the Kucha\v{r} and especially Dirac conceptualizations, it is hard to construct a~suf\/f\/iciently large
set of beables to describe physical theory, in particular for gravitational theory.

Strategies for dealing with the problem of beables include considering each of the Kucha\v{r} and Dirac positions on the
nature of beables to be suf\/f\/icient.
Kucha\v{r} beables can also be viewed as a~potentially useful halfway house in the construction of Dirac beables.
Partial observables as a~problem of observables/beables strategy is along the lines of this problem being held to be
a~misunderstanding of the true nature of observables/beables.
Partial observables are, rather, entities that are measurable but unpredictable by themselves, predictions here
involving rather correlations between more than one such.
On the other hand, another strategy is to use partial observables as intermediates toward obtaining Dirac
observables/beables.

\section{Kucha\v{r} beables}\label{K-B}

\subsection{Further motivation for Kucha\v{r} beables}\label{K-B-Motiv}

1) The Dirac conjecture (Section~\ref{1.1}) is false by e.g.\ a~technically constructed but not physically motivated counterexample
given in~\cite{HT92}: $\text{\sffamily L} =
\exp(y)\dot{x}^2/2$ suf\/f\/ices, with its $p_x = 0$ constraint being f\/irst-class but not associated with a~gauge
symmetry.

2) The conjecture is contested on further grounds by e.g.\ Kucha\v{r}~\cite{K99} and Barbour--Fos\-ter~\cite{BF08};
this is furthermore directly at odds with~\cite{HT92}.
I point out here that this discrepancy is due to~\cite{HT92} allowing for~$t$-dependent canonical transformations,
Can$_t$.
These map reparametrization-invariant actions to non-reparametrization-invariant actions; ${\cal QUAD}$ is then also not
an invariant form under Can$_t$.
On the other hand, one has to presume that Can$_t$ are not licit in Barbour-type relational perspectives, in which
space/conf\/iguration space and timelessness are primary.
Here temporal relationalism (Appendix~\ref{PoT}) is implemented by reparametrization-invariant actions, and the
principles of dynamics is reformulated to suit there being no primary notion of time.
Consequently ${\cal QUAD}$ and ${\cal LIN}_{\text{\sffamily{L}}}$ are qualitatively dif\/ferent types of
entities in this perspective.

3) Kucha\v{r} beables are, moreover, simpler to f\/ind than Dirac ones; Kucha\v{r} was motivated by this rather than~2).

4) A~set of Kucha\v{r} beables can be extended to produce a~set of Dirac beables (see Section~\ref{Dir-Ex}).

One problem of beables strategy is that Kucha\v{r} beables are all~\cite{BF08, Ear01, Ear02,Kouletsis,Kuchar92, Kuchar93, K99, WuthrichTh}.
Finding Kucha\v{r} beables is uncontroversially a~timeless pursuit through its not involving the quadratic Hamiltonian
constraint.
The downside is that a~constraint of this kind remains as a~frozen equation at the quantum level.
Thus one has to concoct some kind of emergent-time or timeless scheme to deal with this (see Appendix~\ref{PoT}).

\subsection{Examples of Kucha\v{r} beables posed}%\label{K-B-Posed}

I denote a~suf\/f\/icient set of Kucha\v{r} beables to describe one's theory by $K_{\text{\sffamily{K}}}$.

{\bf Example 0.} For theories with no linear constraints such as the Jacobi formulation of mechanics or Misner's
minisuperspace~\cite{FileR}, Kucha\v{r} beables are just {\it any} quantities (subject to a~caveat and rephrasing in
Section~\ref{Fa-De}).

{\bf Example 2.} Kucha\v{r} beables for scaled relational particle mechanics obey
\begin{gather}
\boldsymbol{\{}{\cal L}_{i},   K_{\text{\sffamily{K}}} \boldsymbol{\}} \approx 0,
\qquad
\boldsymbol{\{}{\cal P}_i,   K_{\text{\sffamily{K}}} \boldsymbol{\}} \approx 0.
\label{OLi0}
\end{gather}
Also in the elsewise often simpler case of Example~2.b: pure-shape relational particle mecha\-nics~\cite{FileR} (shapes
are relative-angle and ratio of relative separation information)
\begin{gather}
\boldsymbol{\{}{\cal D}, K_{\text{\sffamily{K}}} \boldsymbol{\}} \approx 0.
\label{OD0}
\end{gather}
Here ${\cal D}:= \sum\limits_{I=1}^{N} \underline{q}^I\cdot \underline{p}_I$ is the {\it
zero dilational momentum constraint}.

{\bf Example 1.} Kucha\v{r} beables for electromagnetism obey
\begin{gather*}
\boldsymbol{\{}{\cal G},   K_{\text{\sffamily{K}}} \boldsymbol{\}} \approx 0,
\end{gather*}
and similarly for Yang--Mills theory and the gauge theories that can be associated with each of these.

{\bf Example 3 or 4.} For GR-as-geometrodynamics, Kucha\v{r} beables obey
\begin{gather}
\boldsymbol{\{}{\cal M}_i, K_{\text{\sffamily{K}}} \boldsymbol{\}} \approx 0.
\label{Gdyn-K}
\end{gather}

{\bf Example 5.} For GR in Ashtekar variables, Kucha\v{r} beables obey
\begin{gather}
\boldsymbol{\{}{\cal M}_i,  K_{\text{\sffamily{K}}} \boldsymbol{\}} \approx 0,
\qquad
\boldsymbol{\{}{\cal G}_I,  K_{\text{\sffamily{K}}} \boldsymbol{\}} \approx 0.
\label{AV-K}
\end{gather}

{\bf Counter-example 7.} Kucha\v{r} beables are {\it not} well-def\/ined for supergravity.
This is because the ${\cal S}_A$'s algebraic primality over ${\cal H}$~-- that the commutator of two ${\cal S}_A$
produces ${\cal H}$~\eqref{SSH} but not vice versa~-- means that supergravity's ${\cal LIN}_{\text{\sffamily{L}}}$ do not form a~subalgebraic structure of constraints.
Then by the Casalbuoni brackets version of~\eqref{CCB}, a~consistent notion of beables/observables cannot be associated
with supergravity's ${\cal LIN}_{\text{\sffamily{L}}}$.
Thus the notion of Kucha\v{r} beables is {\it not} available for supergravity, whether as a~problem of beables
resolution in itself or as a~well-def\/ined halfway house in the construction of Dirac beables.

\subsection{Kucha\v{r} beables examples resolved}%\label{K-B-Resolved}

\begin{figure}[ht] \centering
\includegraphics[width=0.70\textwidth]{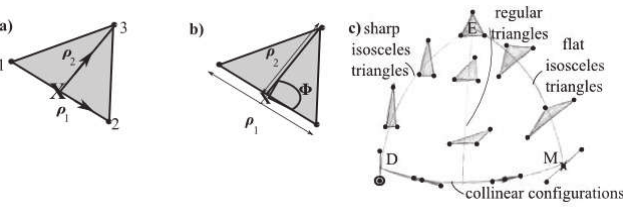}

\caption{The relational triangle.
a)~Relative Jacobi vectors.
X~is the centre of mass of particles~1 and~2.
b)~Their magnitudes, base and median labels, and the angle between them.
c)~The triangleland sphere, with what triangles correspond to which points, is 6 copies of the given 1/3-hemisphere:
Kendall's spherical blackboard~\cite{FileR}.
The 6 copies correspond to dif\/ferent possible labellings of the triangle.
D~is a~double collision, M~is a~merger and E~is the equilateral triangle conf\/iguration.
In comparison with Wheeler's well-known depiction of Superspace($\Sigma$)~\cite{Battelle}, both are clearly spaces of
spaces, but the relational triangle's clearly has the simpler mathematics that renders it of further use as a~model
arena.}\label{Fig-2}
\end{figure}

Example 0 is straightforward to resolve.
Additionally, if linear constraints have been reduced out by whatever means, then one has arrived at a~situation with equal status to Example~0.
This is then the case for Examples~2 and 2.b.
Pure-shape relational particle mechanics is simpler~\cite{QuadI}: classical Kucha\v{r} beables are functionals of the
shapes $\text{S}^{\text{\sffamily{A}}}$ and their conjugate shape momenta~$\text{P}^{\text{S}}_{\text{\sffamily{A}}}$,
\begin{gather*}
K_{\text{\sffamily{K}}} = \big\{\text{F}_{\text{\sffamily{K}}}\big[\text{S}^{\text{\sffamily{A}}}, \text{P}^{\text{S}}_{\text{\sffamily{A}}}\big]\big\}.
\end{gather*}
For scaled relational particle mechanics, classical Kucha\v{r} beables are functionals of a~scale variable~$\sigma$, its
conjugate scale (dilational) momentum $\text{P}_{\sigma}$ and the shape and shape momenta,
\begin{gather*}
K_{\text{\sffamily{K}}} = \big\{\text{F}_{\text{\sffamily{K}}}\big[\text{S}^{\text{\sffamily{A}}}, \sigma, \text{P}^{\text{S}}_{\text{\sffamily{A}}}, \text{P}_{\sigma}\big]\big\}.
\end{gather*}
For example, for the scaled relational triangle~\cite{AHall}, the shape space is the sphere.
Using mass-weighted relative Jacobi vectors $\underline{\rho}_1$, $\underline{\rho}_2$ (Fig.~\ref{Fig-2}a) convenient forms for the
shapes are $\Theta = 2\operatorname{arctan}(\rho_2/\rho_1)$ and $\Phi = \operatorname{arccos}\big({\underline{\rho}}_1 \cdot
{\underline{\rho}}_3 / \rho_1 \rho_3 \big)$.
These are geometrically the spherical polar coordinates on the shape space sphere (Fig.~\ref{Fig-2}c).
The scaled relational triangle also has a~scale variable: the moment of inertia, $I =
\sum\limits_{i=1}^{2}\rho_i^2$, or sometimes more conveniently its square
root~\cite{FileR, LR97}.
The relational triangle's pure-shape momenta are then a~relative angular momentum (between the base and the median)
conjugate to~$\Phi$ and a~relative dilational momentum (dilatation of the base's length relative to the median's length)
The relational triangle's scale momentum is an overall dilatation.

Shape and scale space is $\mathbb{R}^3$ topologically but not metrically (though it is conformally f\/lat).
The corresponding `Cartesian' coordinates are the Hopf--Dragt coordinates~\cite{FileR, LR97} (after the well-known {\it
Hopf map}: $\mathbb{S}^3 \rightarrow \mathbb{S}^2$):
\begin{gather}
{\rm Dra}_1:= 2{\underline{\rho}}_1 \cdot {\underline{\rho}}_2,
\qquad
{\rm Dra}_2:= 2\{{\underline{\rho}}_1
\times  {\underline{\rho}}_2\}_{3},
\qquad
{\rm Dra}_3:= {\rho_2}^2 - {\rho_1}^2.
\label{Dra}
\end{gather}
These and their conjugate momenta $\Pi^{\rm Dra}_i$ are a~useful repackaging of the information in the above scale-shape
split objects.
Then
\begin{gather}
K_{\text{\sffamily{K}}}:= \big\{\text{F}_{\text{\sffamily{K}}}\big[{\rm Dra}^{i}, \Pi^{\rm Dra}_{i}\big]\big\}.
\label{Tri-K}
\end{gather}
See~\cite{QuadI} for their relational quadrilateral counterparts and~\cite{FileR} for the relational~$N$-a-gon
case covered in less detail.

{\bf Example 1.} i)~The electric f\/ield $\underline{E}$ and the magnetic f\/ield $\underline{B}$ are Kucha\v{r} beables for electromagnetism.
ii)~The {\it Wilson loops}
\begin{gather*}
W_{\gamma}[A^i] = \exp \left(i\oint_{\gamma}A_i(y)\textrm{d} y^i \right)
\end{gather*}
(here~$\gamma$ is a~path in space) are also Kucha\v{r} beables for electromagnetism.
Furthermore ii) generalizes to Yang--Mills theory upon introducing tracing over the internal indices.
These loop variables form an overcomplete set of such beables (there are {\it Mandelstam identities} between them); this
point is well-covered in e.g.~\cite{GPbook}.
As regards the signif\/icance of this example, the counterpart of such loops in the Ashtekar variables case are indeed
{\it the} loops in loop quantum gravity.

Formal strategy 1) One can also act with $\mathfrak{g}$ and then perform an operation involving the
whole of $\mathfrak{g}$ (e.g.\
summing, integrating, averaging, taking an inf, sup or extremum) in order to construct formally $\mathfrak{g}$-invariant expressions that serve as Kucha\v{r} beables~\cite{BI}.

Formal strategy~2) In some cases also one knows formally what the $\mathfrak{g}$-invariant expressions
are.
`Formal' here refers to not having a~concrete basis of these such as the above Hopf--Dragt coordinates for triangleland.

{\bf Example 3 or 4.} For GR as geometrodynamics the classical Kucha\v{r} beables are, formally as per strategy~2),
functionals of the 3-geometries and associated momenta,
\begin{gather*}
K_{\text{\sffamily{K}}} = \big\{\text{F}_{\text{\sffamily{K}}}[\mathfrak{G}\text{eom}, \Pi^{\mathfrak{G}\text{eom}}]\big\}.
\end{gather*}
Following strategy 1) instead, one can use entities integrated over all space (but they are not local) or integrated
over Dif\/f($\Sigma$) (but the measure of integration in such expressions remains formal).

{\bf Example 5.} For Ashtekar variables formulations of GR, to commute with ${\cal M}_i$ in addition to with ${\cal G}_I$, one
needs to consider the dif\/feomorphism-invariant classes of loops; this coincides with the mathematical def\/inition of {\it knots}.
Classical Kucha\v{r} beables are then, formally in the sense of strategy~2), functionals of knots and associated
momenta,
\begin{gather*}
K_{\text{\sffamily{K}}} = \{\text{F}_{\text{\sffamily{K}}}[\mathfrak{K}\text{not},
\Pi^{\mathfrak{K}\text{not}}]\}.
\end{gather*}

{\bf Example 6.} The more standard (canonically untransformed) bein presentation of GR involves using the conf\/iguration space
Bein($\Sigma$) in place of Riem($\Sigma$).
Then Dif\/f($\Sigma$) and the local Lorentz transformations are quotiented out in order to pass to a~reformulation of the
information contained in Superspace($\Sigma$).
Nor is this an empty variant of formalism since inclusion of fermions (e.g.\ Einstein--Dirac theory) requires reformulation away from metric variables.

{\bf Example 7.} For supergravity, one cannot just quotient out the linear constraint generated supersymmetric generalization
of Dif\/f($\Sigma$) because of of ${\cal LIN}_{\text{\sffamily{L}}}$ not forming a~subalgebraic structure of constraints.
Thus `SuperSuperspace($\Sigma$)'~-- the na\"{\i}ve supersymmetric generalization of Wheeler's Superspace($\Sigma$) for
GR as geometrodynamics~-- turns out not to be well-def\/ined.
In this case, one has to consider the fully reduced conf\/iguration space and the full notion of Dirac beables as per Section~\ref{D-B}.

{\bf Example 8.} (subcase of geometrodynamics of cosmological relevance).
Kucha\v{r} beables for perturbatively inhomogeneous cosmology about a~homogeneous isotropic $\mathbb{S}^3$
minisuperspace with single scalar f\/ield matter are exposited in~\cite{ABrackets} based on the earlier work
in~\cite{Bardeen,HallHaw1,HallHaw5,HallHaw6, BM,HallHaw4,HallHaw3, HallHaw2}.
These are in terms of a~countable inf\/inity of mode coef\/f\/icients for the small perturbations.
These constitute an explicit $\mathbb{S}^3$ basis much like the Hopf--Dragt coordinates for relational triangle.
This demonstrates how some regimes of GR are simpler and are usefully modelled by relational particle mechanics such as
this review's relational triangle model.

{\bf Example 9.} Other models for which Kucha\v{r} beables are known include a~few midisuperspaces (inhomogeneous but still
nontrivially symmetric models that are more amenable to calculations than fully general models).
For instance, a) some spatially compact without boundary Gowdy models~\cite{TorreGowdy}.
These are once again functions of an inf\/inite number of mode coef\/f\/icients.
b)~Some open\footnote{In Examples 8 and 9, I just give citations to keep this review of manageable length.
Aside from here, I also restrict this review to universes that are spatially compact without boundary.
Asymptotically-f\/lat models have further notions of asymptotic observables/beables and interior observables/beables.
Also far from all open models are asymptotically f\/lat, so the study widens further upon consideration of open models.
Likewise we do not have space in this review to consider the notion of observables/beables in holographic theories.
\label{foo-log}
}   midisuperspace models with known Kucha\v{r} beables are the cylindrical gravitational wave~\cite{Bubble} and
spherically symmetric gravitational models~\cite{Kuchar94}.

\section{Classical Dirac beables}\label{D-B}

\subsection{Motivation for Dirac beables}\label{D-B-Motiv}

Perhaps instead the problem of beables is to be resolved by f\/inding Dirac beables.
These however may be dif\/f\/icult objects to construct in practise.
E.g.\
explicit construction of Dirac beables is subject to the caveat of requiring explicit solution of a~model's
dynamics~\cite{Hoehn, Smolin}, which is in general blocked due to the onset of chaos.
Each of the Kucha\v{r} beables and partial observables positions can be interpreted as a~halfway houses toward
construction of Dirac beables.
The former is clearly by applying one further partial dif\/ferential equations restriction to one's set of Kucha\v{r}
beables: the commutation also with the quadratic constraint.
The latter is via methods developed by Dittrich and Thiemann~\cite{Dittrich1, Dittrich2, Thiemannbook}.

\subsection{Examples of Dirac beables problems posed}%\label{D-B-Posed}

{\bf Example 1.} In electromagnetism, Yang--Mills theory and the gauge theories associated with each, Dirac is
equivalent to Kucha\v{r} for beables, so just take what is said in Section~\ref{K-B} about these theories' beables.
This is also the case for temporally-absolute conf\/igurationally-relational mechanics.

{\bf Example 0.} There is just one condition to be solved for each of spatially absolute mechanics and
minisuperspace:
\begin{gather}
\boldsymbol{\{} {\cal E} ,  D_{\text{\sffamily{D}}} \boldsymbol{\}} \approx 0,
\label{ED=0}
\qquad
\boldsymbol{\{} {\cal H} , D_{\text{\sffamily{D}}} \boldsymbol{\}} \approx 0.
%\label{HD=0}
\end{gather}

{\bf Example 2.} Relational particle mechanics have just one more condition to be solved on top of an already-solved
set of conditions~\eqref{OLi0}, \eqref{OD0} with $K_{\text{\sffamily{K}}} \longrightarrow
D_{\text{\sffamily{D}}}$.
It is schematically also of the form~\eqref{ED=0}.

{\bf Example 3 or 4.} GR as geometrodynamics and in terms of Ashtekar variables both have just one more condition to be solved
on top of a~given set of conditions, the $K_{\text{\sffamily{K}}} \longrightarrow
D_{\text{\sffamily{D}}}$ of~\eqref{Gdyn-K} and~\eqref{AV-K} respectively.

{\bf Example 7.} Having presented a~reason why the problem of f\/inding Kucha\v{r} beables/obser\-vab\-les for supergravity is not
well-def\/ined, I now pose the question of f\/inding a~full set of classical and then quantum Dirac beables/observables for
supergravity.
This in fact appears to be a~new question, just beyond the frontier in~\cite{DEath} of f\/inishing to construct the
classical Dirac brackets algebra for supergravity.

\subsection{Examples of Dirac beables problems solved}\label{Dir-Ex}

For Example 0, i) See e.g.~\cite{ATU93} for a~direct construction of classical Dirac beables for minisuperspace.

ii) Halliwell~\cite{H03} gave a~classical-level construct; for a~simple~$k$-$d$ particle mechanics model and
$\delta^{(k)}$ the~$k$-$d$ delta function, it is of the form
\begin{gather*}
A(\text{{\bf q}}, \text{{\bf q}}_0, \boldsymbol{p}_0) = \int_{-\infty}^{+\infty} \textrm{d} t \, \delta^{(k)}\big(\text{{\bf
q}} - \text{{\bf q}}^{\text{cl}}(t)\big).
\end{gather*}
Here $\text{{\bf q}}^{\text{cl}}(t)$ is a~conf\/iguration space vector valued classical
solution labelled by initial data $\text{{\bf q}}_0$, $\boldsymbol{p}_0$.
It is necessary in this construct to treat the whole path rather than just segments of it.
This is because elsewise the endpoints of segments contribute right-hand-side terms to $\boldsymbol{\{}{\cal
H}, A\boldsymbol{\}}$.
Whilst these Dirac beables are built out of histories, the f\/inal constructs themselves are integrals over all times,~by
which these are indeed beables as opposed to histories beables (see Section~\ref{HB} for these).
This construct extends both to minisuperspace GR~\cite{H03, H09, H10} and to the triangleland relational particle
mechanics~\cite{AHall} subcase of Example~2 formulated in terms of its Kucha\v{r} beables~\eqref{Tri-K}, which provides
a~solution to Example~2.
The latter case involves use of the three Hopf--Dragt coordinates of~\eqref{Dra} in place of the 3-$d$ case of
$\text{{\bf q}}$,   Thus additionally it is an example of building on the halfway house of having constructed a~set of
Kucha\v{r} beables.

For Example 3 or 4, as regards GR beyond minisuperspace, i) see~\cite{ABrackets} for some Dirac beables for the Halliwell--Hawking model.

ii) Dirac beables are sometimes also explicitly known~\cite{TorreGowdy} for some of the Gowdy midisuperspace models.

iv) In outline, Dittrich's~\cite{Dittrich1, Dittrich2} general formal power series expansion objects for GR
are of the form
\begin{gather*}
D_{\phi} = \sum_{n = 0}^{\infty} \frac{1}{n!} \{\text{\sffamily F}\}^n \boldsymbol{\{} \phi , \overline{\cal C} \boldsymbol{\}}_{(n)}.
\end{gather*}
Here~$\phi$ are dynamical f\/ields, $\text{\sffamily F}^{\mu}:= X^{\mu} - Y^{\mu}(X^{\mu})$ is a~gauge-f\/ixing equation for
$Y^{\mu}$ spacetime scalar functions, and $\overline{\cal C}_{\mu}$ are particular linear combinations of the GR
constraints~\cite{HT92}.
Also $\boldsymbol{\{}~,~\boldsymbol{\}}_{(n)}$ is an `$n$ times iterated Poisson bracket',
i.e.\ $n$ Poisson brackets nested inside each other.
Each $\overline{\cal C}_{\mu}$ is contracted with that on one power of ${\text{\sffamily F}}^{\mu}$.
See~\cite{Dittrich1, Dittrich2} for the conceptually relevant points of how this construct 1)~exemplif\/ies proceeding to
Dirac beables via a~partial observables halfway house.
2)~That it involves some partial observable acting as a~clock variable for the others.
See also~\cite{DittTamb2, DittTamb1} for an outline of this perturbative approach in which an Abelian set of constraints
is iteratively produced, alongside the application of this construction to the important case of inhomogeneous
cosmological perturbations.
This approach has already been recently covered in~\cite{Dittrich1, Dittrich2, PSS10, Tambornino, Thiemannbook}, so I
detail it here no further.

The explicit construction problems of Section~\ref{D-B-Motiv} do not af\/fect Halliwell's formal expressions integrated over all time, but
do also apply to Dittrich's power-series construct.

\section{Local observables and beables: fashionables and degradeables}\label{Fa-De}

That coordinates are not in general globally def\/ined on closed conf\/iguration spaces follows from~$\mathbb{S}^2$ being
e.g.\ the shape space for 3 particles in 2-$d$.
Furthermore, classical beables brackets are partial dif\/ferential equation conditions.
Partial dif\/ferential equation solutions seldom form a~globally-coherent whole (e.g.\ they seldom admit global well-posedness or explicit global solutions).
This is far more serious a~mathematical problem than a~mere patching together of coordinates.

{\it Fashionables} are observables that are local in time and space.

{\it Degradeables}, on the other hand are beables that are local in time and space.

These are f\/itting nomenclature for local versions of these concepts along the lines of the expressions `fashionable in
Italy', `fashionable in the 1960's', `degradeable within a~year' and `degradeable outside of the fridge' all making good
sense.
Additionally, fashion is in the eye of the beholder~-- observer-tied, whereas degradeability is a~mere matter of being
rather than of any observing.
Note that the above use in detail of this nomenclature is my own~\cite{AHall}.
Bojowald et al.~\cite{Bojo1, Bojo2, Bojo3, Bojo4} had previously introduced the term `fashionables' without making
distinction between observables and beables, i.e.~their use of the name `fashionables' covers {\it both} of my uses of
`fashionables' and `degradeables'.
Moreover,~\cite{Bojo1, Bojo2, Bojo3, Bojo4} also provide computations for these quantities, which serve for either of
these interpretations, thus also providing a~means of constructing what I term degradeables if one adopts a~realist
interpretation of QM.~\cite{Bojo1, Bojo2, Bojo3, Bojo4} furthermore exemplify patching.
Patching quite clearly ties well with the partial observables approach, though this notion applies also to the
Kucha\v{r} and Dirac conceptualizations of observables or beables.
See Section~\ref{Q-Beables} for further details of Bojowald et al. work at the quantum level.

As further examples, in fact, examples of Section~\ref{K-B} are in general but Kucha\v{r} degradeables rather than globally
well def\/ined Kucha\v{r} beables.

Also, Dittrich's power series construct depends on the `clock variable' conjugate to the constraint being well-def\/ined,
which is in general only local.

Finally, Halliwell's integration over all time construct has the global problem that one's choice of time function does
not generally hold throughout space or for all values of that time.
There is an issue of operational useabilty for entities that require evaluation over all of history.
Non-globality of many emergent and hidden times clashes with how the version of Halliwell's construct based on
integrating over f\/inite time intervals fails to commute with ${\cal H}$.

\section{Dif\/feomorphism-specif\/ic issues with classical beables}\label{Diff}

Passing from ordinary gauge theory to GR substantial increases conceptual and technical complexity.
I begin by recollecting two early no-go results about Dirac beables.

\subsection{Kucha\v{r}'s and Torre's no-go theorems}%\label{No-Go}

{\it Kucha\v{r}'s no-go theorem}~\cite{Kuchar81b}.
Nonlocal objects of the form
\begin{gather}
\int_{\Sigma} \textrm{d}^3x \,{\text{K}}_{ij}(x^k; h_{lm}]  p^{ij}(\underline{{x}})
\label{K1981}
\end{gather}
are not Dirac beables (for ${\text{K}}_{ij}$ some general spatial tensor-valued mixed function-functional).
This result makes use that metric concomitants are in general built out of covariant derivatives of the Riemann tensor.
It then proceeds by proving inductively on the number of covariant derivatives that $\text{K}_{ij}$ cannot contain
concomitants with that number of covariant derivatives, by use of algebraic and integrability arguments.

{\it Torre's no-go theorem}~\cite{Torre93} (see also~\cite{+Torre2, Carlip01,+Torre1}).
Local functionals
\begin{gather*}
\text{T}(x^i; h_{kl}, p^{nm}]
\end{gather*}
are not Dirac beables either.
This uses that local observables correspond to local `hidden symmetry' but that the latter's cohomological
classif\/ication then leaves no viable options.

\subsection{Interpretations of dif\/feomorphisms}%\label{Diff-Int}

For $\mathfrak{s}$ some dif\/ferential manifold, I make standard use of Dif\/f($\mathfrak{s}$) for the dif\/feomorphisms (usually actively interpreted).
Dif\/f($\mathfrak{m}$) and Dif\/f($\Sigma$) are then cases of particular relevance.
\begin{gather*}
\text{Dif\/f}(\mathfrak{m}) = \{\epsilon(X^{\mu})\}
\end{gather*}
corresponding to coordinate transformations\footnote{Inf\/initesimal transformations   $X^{\mu} \rightarrow
\widetilde{X}^{\mu}$ can be written as $X^{\mu} - \widetilde{X}^{\mu} = \epsilon^{\mu}$.
Viewed as solutions in terms of phase space variables, the right hand side functions here are so-called {\it
descriptors} (a fairly standard gauge-theoretic notion, see e.g.~\cite{A67, BK72}).
For GR, descriptors are arbitrary functions of $X^{\mu}$, $h_{ij}$ but not of~$N$ or $N^i$.
Section~\ref{Dir-Ex}'s~$\text{\sffamily{F}}^{\mu}$, the below $\xi^{\mu}$ and Section~\ref{Weyl}'s Weyl scalars can
each be viewed as particular uses of descriptors.
\label{desc}}   $X^{\mu} \rightarrow \widetilde{X}^{\mu} = f^{\mu}(X^{\nu})$.
However GR is invariant under a~larger group~\cite{BK72}: the {\it diffeomorphism-induced gauge group},
\begin{gather*}
\text{Digg}(\mathfrak{m}):= \big\{\epsilon\big(X^{\mu}; \phi^{\Gamma}(X^{\mu})\big]\big\}.
\end{gather*}
Here $\phi^{\Gamma}(X^{\mu})$ denote the f\/ields in one's theory (metric $g_{\mu\nu}(X^{\mu})$ and matter f\/ields).
Digg($\mathfrak{m}$) might also be denoted BK($\mathfrak{m}$) after Bergmann and
Komar~\cite{BK72}, though they themselves referred to it as the `$Q$-group', and Pons, Salisbury and Sundermeyer prefer
to use the Bergmann--Komar name for the f\/inal group in this subsection.
Incidentally, the existence of this larger invariance does not by itself dictate that it is {\it the} gauge group.
One can argue rather for freedom to choose dif\/ferent suitably compatible groups to be physically irrelevant, and then
consider the outcome of each theory~\cite{AM13}.
Inconsistencies and observational unviability then serve to kill of\/f choices not realized in nature.
Adopting Digg($\mathfrak{m}$) invariance feeds into one type of resolution of the frozen formalism
problem (see Appendix~\ref{PoT}), so there are theoretical reasons for considering this group.

Next,
\begin{gather*}
\text{Data}(\mathfrak{m}):= \big\{\epsilon\big(X^{\mu}; \phi^{\Gamma}_{\text{C}\text{D}}(X^{\mu})\big]\big\}.
\end{gather*}
Here `CD' denotes `depends on the f\/ields only via the {\it Cauchy data} on a~spatial hypersurface~$\Sigma$.
This notion was adopted by Dirac, and so Bergmann--Komar called it `D-group'.
The associated invariance concerns transformations that are unchanged under 4-$d$ coordinate transformations that reduce
to it on initial~$\Sigma$ on which the canonical Cauchy data are def\/ined.
E.g.\ any spatial 3-vector, $h_{ab}$ or $p^{ab}$ are Data-invariant.
Dirac adopted this since his application of Poisson brackets is appropriate to Data($\mathfrak{m}$) but
not to Dif\/f($\mathfrak{m}$) or Digg($\mathfrak{m}$).

I denote the f\/inal group of particular interest by PDigg($\mathfrak{m}$): the projected
version of Digg($\mathfrak{m}$).
Pons, Salisbury and Sundermeyer~\cite{PSS10} refer to this as the Bergmann--Komar group, since it is also given
in~\cite{BK72}.
However, it is surely more natural in talking about the two new groups in~\cite{BK72} to name this one by its
distinctive feature of being a~projection.
This is as opposed to trying to f\/ind a~less clear distinctive feature that refers to the larger group.
I note moreover that Bergmann and Komar themselves did not themselves know about this group's geometrical interpretation
in terms of projection.
This was, rather, later elucidated by Pons, Salisbury and Shepley~\cite{PSS97}.
Before considering this interpretation, it is worth pointing out the form of the group:
\begin{gather}
\text{PDigg}(\mathfrak{m}) = \big\{\epsilon\big(X^{\mu}; \phi^{\Gamma}(X^{\mu})\big] \in \text{Digg}(\mathfrak{m}) \,\big| \,
\epsilon = n^{\mu}(X^{\nu})\xi^0 - \delta^{\mu}_a\xi^a\big\}.
\label{Bolt}
\end{gather}
Here, the $\xi^{\mu}$ are descriptors (see footnote~\ref{desc}) of the particular form $\xi^{\mu}(X^{\mu};
\phi^{\text{\sffamily{P}}}(X^{\mu})]$.
$\phi^{\text{\sffamily{P}}}$ denotes a~set of metric and matter f\/ields that now specif\/ically exclude the GR
lapse and shift.
Finally, ${\cal H}_{\mu}$ denotes the 4-vector of constraints $[{\cal H}, {\cal M}_i]$.
Then the projecting in question is from conf\/iguration--velocity space to phase space, i.e.~associated with a~Legendre
map.
It ef\/fectively means that the induced gauge group for $Q^{\text{\sffamily{A}}}$,
$P_{\text{\sffamily{A}}}$ is smaller than that for $Q^{\text{\sffamily{A}}}$,
$\dot{Q}_{\text{\sffamily{A}}}$.
Thus PDigg's `$\text{P}$' can be taken to stand for `phase' as well as for `projection'.
In ef\/fect, Digg($\mathfrak{m}$) itself cannot be completely realized in phase space (and
Dif\/f($\mathfrak{m}$) is not meaningful in this sense either); this is what motivates adopting
PDigg($\mathfrak{m}$) instead.
The corresponding active canonical phase space transformation is (see Fig.~\ref{Fig-1} for the meaning of~$P_{\mu}$)
\begin{gather*}
G_{\xi} = P_{\mu}\dot{\xi}^{\mu} + \{{\cal H}_{\mu} + N^{\rho}f^{\nu}{}_{\mu\rho}P_{\nu}\}\xi^{\mu}.
\end{gather*}
Also here, $f^{\nu}{}_{\mu\rho}(h_{ij})$ are the Dirac algebroid's structure functions wrapped up in the spacetime
tensor form corresponding to expressing the constraints as a~spacetime vector ${\cal H}_{\mu}$.

Bergmann and Komar~\cite{BK72} also posited a~number of relations between the groups involved.
This led them to conclude that Dirac and Bergmann beables turn out to coincide, but they provide no proofs for the
underlying relations and this claim has since been contested~\cite{LP04, WuthrichTh}.

\subsection{The dif\/ference between a~Hamiltonian and a~gauge generator}%\label{Ham-Gau}

The next two subsections are based on the `scholium' presented in dialogue form~\cite[pp.~21--22]{PSS09}, and in Section~3.3
of~\cite{PSS10}.

The `evolution' generator $\delta t \{N^{\mu}H_{\mu} + \dot{N}^{\mu}P_{\mu}\}$ does serve to replace solutions at
time~$t$ by the original solutions evaluated at $t - \textrm{d} t$.
However, this is merely its action on one particular member of each equivalence class of solutions.
I.e.\
the particular member for which the lapse and shift form the chosen explicit function $N^{\mu}$ as per Fig.~\ref{Fig-1}.
Its action on all other members of these equivalence classes generates variations dif\/ferent from global time
translations.

In more detail, points $p\in S$ = \{the space of the dynamical f\/ields~$\phi$\} are specif\/ic spacetimes plus
matter f\/ields when relevant: solutions of the equations of motion as described in a~particular coordinatization.
Also use $\text{D}$ to denote the data for the dynamical f\/ields on some spatial hypersurface that is labelled by
`initial time'~$t_0$.
Given a~specif\/ic selection of the arbitrary functions of the dynamical variables $\lambda^{\mu}$, the corresponding
Dirac Hamiltonian is $\text{\sffamily H}(t) = N^{\mu}{\cal H}_{\mu} + \lambda^{\mu} P_{\mu}$.
This dictates~-- via the Poisson brackets~-- the time evolution in~$p$.
In particular, for an inf\/initesimal $\textrm{d} t$, this Hamiltonian gives what the f\/ield data $\text{D}^{\prime}$ are
on the subsequent spatial hypersurface labelled by $t_0 + \delta t$.
If we carry out this procedure for all times~$t$, it of course results in a~`null operation': we have remained exactly
at the same point $p \in S$.
This simply ref\/lects that the dynamics as described by a~given observer takes place within a~given spacetime in a~given
coordinatization.

Next consider the gauge generator that, after suitable choice of the descriptors, happens to coincide in its
mathematical form with the Dirac Hamiltonian at time~$t_0$.
By this coincidence, its action likewise transforms the f\/ield data~$\text{D}$ into $\text{D}^{\prime}$.
However, these data $\text{D}^{\prime}$ are now to be interpreted at time~$t_0$, since the notion of gauge
transformations in question are {\it equal-time} actions.
What has occurred is that we have moved from~$p$ to another, albeit gauge-equivalent, spacetime~$p^{\prime}$.
(I.e.\ it is mathematically another point in~$S$, but it corresponds to the same physics.)   Then suppose we undertake the same
procedure for any time~$t$ whilst continuing to assume that the descriptors at time~$t$ match up with the lapse and
shift at~$t$.
Then we end up having mapped the whole spacetime~$p$ to $p^{\prime}$.
Notice that the f\/ield conf\/igurations in~$p$ and $p^{\prime}$ dif\/fer solely as regards their time labels.
Thus a~passive dif\/feomorphism $t \rightarrow t - \delta t$ renders both descriptions mathematically identical.
This demonstrates that the gauge generator's capacity to mimic the Hamiltonian is {\it conceptually unrelated} to there
being real physical evolution in a~given spacetime~$p$.
Thus dynamical evolution in~$p$ is not the same thing as gauge action on~$p$.

\subsection{The `nothing happens' fallacy}\label{Fall}

`\eqref{Fally} means that nothing happens' is a~common type of frozen argument (Appendix~\ref{PoT}~discusses others).
However, the inference that `nothing happens' is a~fallacy on the following grounds~\cite{PSS09}.

On the one hand, following from the preceding subsection, we have an `evolved conf\/igura\-tion'~$\text{D}^{\prime}$ lying to
the future of of an `initial conf\/iguration'~$\text{D}$.
On the other hand, $\text{D}$ and $\text{D}^{\prime}$ are related by a~gauge transformation.
Since `gauge transformations do not change the physics', we deduce that `the physics' in $\text{D}$ and
$\text{D}^{\prime}$ are the same.
So the future conf\/iguration is gauge-equivalent to the initial conf\/iguration and therefore `nothing happens'.
The fallacy comes from each of the two hands using a~single common language for two sets of things that are in fact
conceptually dif\/ferent in each case.
Recollect from Section~\ref{1.3} that there are two notions of gauge transformation: Dirac's and Bergmann's.
There are furthermore also two corresponding notions of `the physics'.
The second hand involves mapping solutions of the equations of motion to other such solutions, and so requires the
`entire f\/ield conf\/igurations' of a~`whole-path', `whole-history' or `whole-spacetime' physics perspective, and thus
involves Bergmann's notion of gauge.
In contrast, the f\/irst hand involves `conf\/igurations at a~given time $t_0$' ($\text{D}$ and $\text{D}^{\prime}$), i.e.\ a~`time-sliced' physics' dynamical perspective, and thus involves Dirac's notion of gauge.
Thus the two hands in fact use both distinct notions of `gauge' and corresponding distinct notions of `the physics'.
Since the `nothing happens' argument does not take these dif\/ferences into account, it is rendered fallacious.
See~\cite{JP09,PS05, PSS09b, Thiemannbook} for further support of this point.
Finally, this resolution of the `nothing happens paradox' corresponds to the distinction between time-dependent beables
$D_{\text{\sffamily{D}}}(t)$ for~$t$ an intrinsic coordinate scalar that constitutes a~gauge f\/ixing, as
opposed to just a~constant $D_{\text{\sffamily{D}}}$ as occurred in the opening paragraph (see Section~3.4
of~\cite{PSS10} for more on this point).
Clearly the former are not `constants of the motion'!

\subsection{Using Weyl scalars as observables/beables}\label{Weyl}

The Weyl scalars are {\it spacetime} scalars; they are built from the Weyl tensor irreducible part of the spacetime
Riemann tensor.
See~\cite{DG} for their detailed forms (not needed for this review's discussion).
The Weyl scalars are of value as per below, as well as being foundational for the Newman--Penrose formulation of
GR~\cite{PR,PR+}.

The Weyl scalars can additionally be considered as a~concrete proposal~\cite{BK72,PS05} for observables/beables in the
sense of Bergmann.
Indeed Bergmann and Komar\footnote{Note that   while this subsection involves the same authors as the previous four subsections it
concerns a~largely disjoint idea.}   converted the spatial components of the spacetime Riemann tensor and
contractions with the spatial hypersurface normal $n^{\mu}$ to be purely in terms of canonical variables (meaning in
this case $h_{ab}$, $p^{ab}$ and {\it not}~$N$ or $N^i$).
Thus the Weyl scalars can be written in terms of the canonical variables, as bef\/its many of the expectations about
observables/beables.
In this application, they are to be interpreted as intrinsic coordinates, and also as `making use of a~set of scalars as
a~gauge f\/ixing'.
Finally note that these scalars can in principle be observed locally and in a~convenient manner, e.g.\ by use of a~Szekeres gravitational compass~\cite{Szekeres}.

\subsection{Comments}%\label{No-Go-2}

{\bf Revisiting Torre's no-go theorem.}
\cite{PS05} posits that Torre proves nonexistence of constant-in-time observables = constants of the motion, built as
spatial integrals.
This is as opposed to dealing with {\it Bergmann} observables such as the Weyl scalars.
On the other hand, Dittrich and Thiemann's approach~\cite{Dittrich1, Dittrich2, Thiemannbook} gets round the Torre no-go
by~\cite{PSS09,PSS10} involving series of Cauchy data derivatives that are in principle up to inf\/inite order.
Finally, Halliwell's classical construct avoids Torre's by not being local in space or time and it avoids
Kucha\v{r}'s by not being of form~\eqref{K1981}.

{\bf Connection with partial observables approaches.}
\cite{PS05} is supportive as is~\cite{PSS09} as per above, and~\cite{PSS09b}.
\cite{PSS09} does limit support in the sense of insisting partial observables be spacetime scalars (a standard tenet of
internal time approaches).
However~\cite{PS05} argues for Weyl scalars exemplifying partial observables (which of course {\it do} in this case
comply with being spacetime scalars).
Thus Dittrich and Thiemann's works on observables/beables has wider conceptual support/motivation than is often
acknowledged.
I.e.\ these avoid the Kucha\v{r} and Torre no-go results and are aligned with the works of Pons, Salisbury, Shepley and
Sundermeyer and preceding works by Bergmann and Komar.

{\bf Reparametrization-invariant model counterpart.}
Some of this section's issues already have nontrivial counterparts for RI theories.
In place of Dif\/f($\mathfrak{m}$), one has
\begin{gather*}
\text{(reparametrizations)}
\qquad
\text{R} = \{\epsilon(t)\},
\end{gather*}
and in place of in place of Digg($\mathfrak{m}$), one has
\begin{gather*}
\text{Rigg} = \{\epsilon(t; q^i(t)]\}
\end{gather*}
(reparametrization-induced gauge group).
Finally, in place of PDigg($\mathfrak{m}$), one has
\begin{gather*}
\text{PRigg} = \big\{\epsilon(t; q^i(t)] \in \text{Rigg} \, | \, \epsilon = \xi(t; q^i(t)]/N\big\}.
\end{gather*}
(That is obtained by applying the formula for $n^{\mu}$ in Fig.~\ref{Fig-1} to~\eqref{Bolt} and truncating to 1-$d$.)   How
complete these reparametrization invariant counterparts are as a~model of dif\/feomorphism-induced groups remains to be
worked out in detail.

Also note that Lee--Wald's tie between Digg($\mathfrak{m}$) nontriviality and GR algebroid
fails to work for Rigg, limiting the extent to which Rigg functions as a~toy model for Digg($\mathfrak{m}$).

{\bf Pitts' work.} I f\/inally mention to keep an eye on Pitts' concurrently produced work that follows in part from the work from Bergmann
through to Pons et al., starting with~\cite{Pitts}.

\section{Beables at the quantum level}\label{Q-Beables}

The simplest notion of QM observables concerns self-adjoint operators which therefore have real eigenvalues and to that
extent are realistic.
This is prior to gauge-theoretic considerations, which are a~further restriction on such operators.

{\bf Strategy 1.} One has classical beables beforehand, one can attempt to promote them to quantum ones.
This might occur at the level of kinematical quantization~\cite{I84} or be viewed as a~process in addition to that.
In either case, one f\/inds that one needs to select a~classical subalgebra of objects to promote to quantum operators.
Perfectly good classical beables can fail to be quantum beables.
This parallels how perfectly good classical symmetries fail to be quantum symmetries due to anomalies arising, in the
sense that both are bracket obstructions upon passing from classical to quantum brackets.
Thus one might term the above phenomenon a~`beables anomaly'.
Schematically, for whichever appropriate pairing of classical and quantum types of bracket,
\begin{gather*}
\boldsymbol{\{} B, {\cal C}_{\text{\sffamily{C}}}\boldsymbol{\}}\, \text{`='}\, 0
\quad
\not\Rightarrow
\quad
\boldsymbol{[}\widehat{B},
\widehat{\cal C}_{\text{\sffamily{C}}}\boldsymbol{]}\Psi = 0.
\end{gather*}
Whether this occurs is moreover dependent {\it twice over} on operator-ordering ambiguities (in the beable operators and
in the constraint operators).
The outcome of the preceding furthermore heavily depends on choice of operator ordering assumed.
Additionally, trying to promote known classical beables to quantum ones also falls afoul of the multiple choice problem
(see Appendix~\ref{PoT}).

{\bf Strategy 2.} One might also start afresh in the quest to f\/ind observables/beables at the quantum level.
This makes particular sense upon realizing that in general the classical and quantum brackets correspond to dif\/ferent
algebras due to global ef\/fects entering at the quantum level~\cite{I84}.
In general, the entities that commuted with the classical constraints with respect to one brackets structure should not
be expected to result in quantum operators that commute with the quantum constraints with respect to an
algebraically-distinct brackets structure!

An issue af\/fecting both strategies in the case of loop quantum gravity is that one has yet to attain a~satisfactory form
for the quantum Hamiltonian constraint, which would then be required as an entity that enters the def\/inition of Dirac
beables.
Another issue af\/fecting both strategies for any suf\/f\/iciently general theory is as follows.
Whereas strategy~1 directly follows on from what is in general a~practical impossibility~-- general classical solution
of equations as complicated as the Einstein f\/ield equations of GR~-- it is hard to believe that strategy~2's starting
afresh at the quantum level would simplify this problem.
Indeed, Dirac~\cite{Dirac} commented that QM seldom simplif\/ies situations that are already complicated at the classical
level.

The {\it quantum problem of beables} is then that, whichever of the above strategies one employs, it is hard to come up
with a~suf\/f\/icient set of these for quantum-gravitational physics.

\subsection{Resolution of Heisenberg picture form of frozen formalism problem}\label{Hall-1}

Adopting Section~\ref{Fall}'s interpretation of the `nothing happens paradox' entails resolution of the quantum frozen
formalism problem in its Heisenberg picture form.
If this problem is resolved in the Heisenberg picture, it would often also be argued that it is resolved in the
Schr\"{o}dinger picture due to these two pictures standardly being unitarily equivalent.
On the other hand, Rovelli would appear to dif\/fer in this regard~\cite{Rovellibook}, by giving the Heisenberg picture
a~privileged status in quantum gravity.

The existence of a~Heisenberg resolution furthermore demotivates timeless approaches such as the na\"{\i}ve
Schr\"{o}dinger interpretation~\cite{NSI1, NSI2}, conditional probabilities interpretation~\cite{PW83}, Page's
approach~\cite{Page1, Page2}, Gambini--Porto--Pullin's~\cite{GPP1, GPP2}, Barbour's~\cite{EOT, B94II} and other forms of
fully timeless records theory~\cite{ARec1, ARec2} (see Appendix~\ref{PoT} for this terminology).

On the other hand, I argued that histories theory (also see Appendix~\ref{PoT}) is distinct in this regard, as are `time from
change' approaches~\cite{ARel2, B94I, Rovellibook}.
There are possibly additionally positions intermediate between the previous two sentences'.
E.g.\
the well-known path integral approaches and `records within histories' approaches~\cite{ARec1, AHall, ARec2,GMH, H03, H99,
H09, H10}.

Finally note that such a~frozen formalism problem resolution does not amount to a~resolution of the quantum problem of
beables itself.

\subsection{Examples of quantum beables}\label{Hall-2}

Quantum Dirac beables obey~\eqref{QDB}, where all second-class constraints have been priorly dealt with at the classical
level by Dirac brackets or extension to ef\/fective phase space.
On the other hand, quantum Kucha\v{r} beables obey~\eqref{QKB}, and are applicable in cases for which the $\widehat{\cal
LIN}_{\text{\sffamily{L}}}$ form a~closed algebraic structure under quantum commutators.

{\bf Examples 0 and 2.} In the absense of any linear constraints, Kucha\v{r} beables remain trivial at the quantum level.
\cite{ATU93} also gives some explicit examples of minisuperspace quantum Dirac beables.
Explicit quantum Kucha\v{r} beables for the relational triangle follow from adopting the Hopf--Dragt coordinates, their
conjugate momenta and an ${\rm SO}(3)$'s worth of shape momenta as the kinematical quantization for that model~\cite{AHall}.

On the other hand, some general features of indirect constructs are as follows.

{\bf Universal Example 1.} Any operator $\widehat{O}$ can be inserted into the construction
\begin{gather}
\widehat{O}_{\text{D}}:= \int \textrm{d} t \, \exp(iH_0 t)   \widehat{O}   \exp(-iH_0 t)
\label{construct}
\end{gather}
for a~suitable notion of time (e.g.\ label time~$\lambda$ in minisuperspace or Newtonian time in mechanics),
which again needs to run over all values of time rather than just some interval.
Formal f\/ield-theoretic generalizations of this construct are also straightforward.
DeWitt's~\cite{DeWitt62} early treatment of construction~\eqref{construct} further specialized to the semiclassical
case.
Marolf~\cite{Marolf09} then treated such objects in the case of QM at the perturbative level.

{\bf Example 3.} Another means of attaining observables/beables in the case of GR, exposited by Giddings, Marolf and
Hartle, involves integrating $\widehat{O}$ over all of spacetime~\cite{G-M-H}.
These authors consider attaining locality by smearing with delta functions.

{\bf Universal Example 2.} Halliwell's semiclassical Dirac beables construction for theory with no linear constraints consists
of the {\it class functional}\footnote{Here,   $\text{{\bf q}}^{\text{cl}}(t)$ is the
classical trajectory, $\text{{\bf q}}_0$, $\boldsymbol{p}_0$ is initial data,~$\theta$ is the step function,
$f_{\text{R}}$ is the characteristic function of region~R, $\varepsilon$~is a~small real number, $S(\text{{\bf
q}}_{\text{f}}, \text{{\bf q}}_0)$ is the classical action between $\text{{\bf q}}_{\text{f}}$
and $\text{{\bf q}}_0$.
See~\cite{HT02} for the detailed form of the prefactor function~$P$.}
\begin{gather*}
\widehat{C}_{\text{R}}[\text{{\bf q}}_{\text{f}}, \text{{\bf q}}_0]:= \theta
\left(\int_{-\infty}^{\infty} \textrm{d} t \, f_{\text{R}}(\text{{\bf q}}_{\text{f}}(t)) -
\varepsilon \right) P(\text{{\bf q}}_{\text{f}}, \text{{\bf q}}_0) \, \exp(iS(\text{{\bf
q}}_{\text{f}}, \text{{\bf q}}_0)).
%\label{gyr}
\end{gather*}
Once again, the specif\/ic example is mechanics but has also been considered for minisuperspace~\cite{H03} and
triangleland relational particle mechanics~\cite{AHall}.
For the f\/irst two cases, it is a~$S$-matrix construct since there are no linear constraints; in the last case it is
a~nontrivial Dirac beables construct since this case has linear constraints to overcome too.
Note that this is an example of `starting afresh' with a~new structure rather than of trying to promote Halliwell's
distinct classical construct to the quantum level.
The class functional can then be used to re-express the decoherence functional between pairs of histories~$\eta$,
$\eta^{\prime}$\footnote{Here $\mathbb{D}\text{{\bf q}}$   is a~measure and is a~$\rho$ a~density matrix.}
\begin{gather*}
{\cal D}\text{e}\text{c}[\eta, \eta^{\prime}] = \int_{\eta}\mathbb{D} \text{{\bf q}} \int_{\eta^{\prime}} \mathbb{D}
\text{{\bf q}}^{\prime} \exp(i\{S[\text{{\bf q}}(t)] - S[\text{{\bf
q}}^{\prime}(t)]\} \rho (\text{{\bf q}}_{0}, \text{{\bf q}}_{0}^{\prime})
\\
\phantom{{\cal D}\text{e}\text{c}[\eta, \eta^{\prime}]}{}
= \iiint \mathbb{D}\text{{\bf q}}_{\text{f}}\mathbb{D}\text{{\bf q}}_{0} \mathbb{D}\text{{\bf
q}}^{\prime}_{\text{f}}   {\widehat{\cal C}}_{\eta}[\text{{\bf q}}_{\text{f}}, \text{{\bf
q}}_{0}]   {\widehat{\cal C}}_{\eta^{\prime}}[\text{{\bf q}}^{\prime}_{\text{f}}, \text{{\bf
q}}^{\prime}_{0}] \Psi(\text{{\bf q}}_{0}) \Psi(\text{{\bf q}}^{\prime}_{0}).
%\label{dunno}
\end{gather*}
Note that the semiclassical Halliwell and quantum Marolf constructs indeed provide examples of beables rather than of
histories beables (a notion along the lines of Section~\ref{HB}).
In each case, this is due to the history content being integrated out by the integration over all~$t$.
Moreover, by involving a~$t$-integral, Halliwell's semiclassical object is not local in time, which, as for the
classical counterpart, would however be a~desirable property in a~beable that can be used in practise.
As in the classical counterpart, the above object manages to commute with $\widehat{\cal H}$ only by having its
integrals run over all time, which is often incompatible in practise with the global nonexistence of e.g.\ emergent and hidden timefunctions (Appendix~\ref{PoT}).
On the other hand, Anastopoulos' distinct histories-based construct~\cite{Ana01} does not have to involve the whole real
line.
The end-product of this is, however, a~histories beable, whereas Halliwell's construct returns in the end just a~beable.

\looseness=-1
A further concern here is whether dif\/f\/iculties ensue if one subsequently forms functionals out of the Kucha\v{r} beables
(specif\/ically, class functionals).
That provides commutation with ${\cal QUAD}$, but now a~functional of beables is not necessarily an beable compatible
with one's previously determined beables.
Thus the multiple choice problem (Appendix~\ref{PoT}) is looming as a~reason for breakdown of the Halliwell procedure being
applied to produce nontrivially-Kucha\v{r} Dirac observables.
And yet, ${\cal LIN}_{\text{\sffamily{L}}}$ is absent by this stage.
Thus it is not $\boldsymbol{[} \widehat{\cal LIN}_{\text{\sffamily{L}}} ,
\widehat{K}_{\text{\sffamily{K}}} \boldsymbol{]}\Psi = 0$ failing to imply that $\boldsymbol{[} \widehat{\cal
LIN}_{\text{\sffamily{L}}} ,  \widehat{{\cal F}(\underline{{x}}; K_{\text{\sffamily{K}}}]} \boldsymbol{]}\Psi = 0$ also holds.
Rather it is an issue of whether the class functionals furnish a~compatible set {\it among themselves} if they are to
now be regarded as beables.

Note also that quantization complicates status of beables via the subalgebra selection criterion (and its own associated
multiple choice problem).
Are the quantum Dirac beables {\it well-defined} as functionals of the quantum Kucha\v{r} beables?

Additionally, note that Halliwell's 2009 version of class functionals~\cite{H09} is more complicated than the above
treatment but succeeds in additionally avoiding the additional manifestation of frozenness that is the `watched kettles
never boil' quantum zeno problem.

Does using Halliwell's more intricate histories-theoretic machinery have any advantages over DeWitt's and
Marolf's constructs?   The answer is yes at the level of handling whole-universe issues and problem of time approaches,
but whether there is practical equivalence between each approach's type of objects remains uninvestigated.

As a~f\/inal example, consider Bojowald et al.~\cite{Bojo1, Bojo2, Bojo3, Bojo4} fashionables approach at the
semiclassical level.
Here, use is made of a~moments expansion to bypass the inner product problem.
The fashionables are real-valued, but solving the constraints gives that this approach's notion of time goes complex
around the semiclassical regime's turning points.
Moreover, in this approach a~time variable's imaginary part becoming signif\/icant is a~diagnostic for that time variable
ceasing to be a~`good clock', due to the onset of non-unitarities in the evolution brought about by the global problem
of time (Appendix~\ref{appendixA}).

\section{Various further brackets and corresponding notions of beables}\label{Var-Bra}

\subsection{Multisymplectic approach}%\label{Symp}

In this approach (see e.g.~\cite{Multisymplectic}), the $\partial/\partial t$ in the standard def\/inition of
momentum~\eqref{mom-vel} is extended to a~$\partial/\partial x^{\text{\sffamily{X}}}$ for $\text{\sffamily
X}$ taking $m > 1$ values.
Then the conventional Hamiltonian curve of the standard $\partial/\partial t$ case, associated with the one tangent
vector, is replaced by a~multivector notion corresponding to~$m$ tangent vectors.
The standard case's Poisson bracket's corresponding symplectic form is now replaced by a~multisymplectic form.
The notion of bracket associated with this is the {\it Schouten--Nijenhuis bracket}: a~{\it graded} Lie bracket on the
multivector f\/ields.

The corresponding notion of observables/beables has been considered by H\'{e}lein and Kouneiher~\cite{Multisymplectic}
(for now in a~classical, non-whole universe setting in which the observables/beables distinction is not signif\/icant).

I f\/inally comment that the multisymplectic approach may have dif\/f\/iculties via putting time and space on even more of an
equivalent footing than is usual.
Whilst this is in part motivated by a~desire to reformulate or replace canonical formulations with manifestly covariant
ones, one also has to bear in mind that time and space are conceptually dif\/ferent entities.

\subsection{Histories brackets and histories beables}\label{HB}

Histories theory is motivated as per Appendix~\ref{PoT} and as regards providing a~more natural interpretation for quantum
cosmology and a~format within which decoherence applies.
All of these issues additionally f\/it in well with realist interpretations of QM and of quantum cosmology.
Moreover some versions of histories theory~\cite{IL, Kouletsis, Savvidou04, Savvidou04b} can be viewed as a~distinct
approach to covariantizing the canonical formulation.
In these versions, conf\/igurations $Q^{\text{\sffamily{A}}}$ have been replaced by whole paths of
conf\/igurations~-- classical histories $Q^{\text{\sffamily{A}}}(\lambda)$~-- in the role of basic canonical
entities.
Here~$\lambda$ is a~label time, though it can be replaced with other notions of time (Appendix~\ref{PoT}) without issue.
The latter are now associated with histories momenta $P_{\text{\sffamily{A}}}(\lambda)$, with {\it histories
brackets} $\boldsymbol{\{}~,~\boldsymbol{\}}_{\boldsymbol{H}}$ holding between the two.
Thus the entirety of the phase space structure has been replaced by a~histories phase space structure.
This is an intriguing possibility, though it is not yet clear whether the study of dynamics based on conf\/igurations
contains any important ingredients that are lost with this paradigm shift to histories being the basic canonical
entities.
In this approach, additionally, the standard notion of constraints is replaced by that of {\it histories constraints},
${\cal C}^{\lambda}_{\text{\sffamily{C}}} = {\cal C}(\lambda)_{\text{\sffamily{C}}}$.
There are then obvious generalizations of the notions of f\/irst- and second-class constraints, Dirac brackets, extended
phase space, ef\/fective constraints and constraint algebra, all now in terms of the histories bracket.

{\it Histories beables}~\cite{Ana01, Kouletsis} are then histories quantities
$B^{\lambda}_{\text{\sffamily{B}}} = B(\lambda)_{\text{\sffamily{B}}}$ that histories-brackets
commute~\cite{IL, Savvidou04} with the histories constraints.
In particular, in a~theory with only f\/irst-class constraints, there is a~notion of Dirac histories beables that
histories-brackets commute with all of these histories brackets constraints,
\begin{gather*}
\boldsymbol{\{} D_{\text{H}}^{\lambda} ,  {\cal C}_{\text{\sffamily{C}}}^{\lambda}\boldsymbol{ \}}_{\boldsymbol{H}} \approx 0.
%\label{HDirObs}
\end{gather*}
In cases in which second-class histories constraints are initially present, this requires using the histories version of
the Dirac bracket or reformulation in terms of histories ef\/fective constraints.
Moreover, some classical theories contain the linear histories constraints ${\cal LIN}_{\text{\sffamily{L}}}^{\lambda}$ as a~subalgebraic structure, alongside a~quadratic histories constraint ${\cal QUAD}^{\lambda}$.
In these cases, a~notion of histories Kucha\v{r} beables is also available:
\begin{gather*}
\boldsymbol{\{} K_{\text{H}}^{\lambda},  {\cal LIN}_{\text{\sffamily{L}}}^{\lambda}\boldsymbol{\}}_{\boldsymbol{H}} \approx 0.
%\label{HKObs}
\end{gather*}

{\bf Note 1.} Despite both being based on classical paths, the Bergmann and histories notions of beables are technically and
conceptually distinct extensions of which gauge groups to attribute to one's physical theory.
This is clear from the shift in basic canonical entities in the latter, with the ensuing introduction of a~histories
brackets structure absent in Bergmann's work and with a~number of subsequent parallels to Dirac's notion of beables.
Nonetheless, Savvidou~\cite{Savvidou04b} showed that histories phase space can also be considered to carry
representations of $\text{Digg}(\mathfrak{m})$.

{\bf Note 2.} The above is a~classical precursor for the quantum {\it histories projection operator} (HPO)
approach~\cite{IL, Savvidou04}.
At the quantum level, histories are paths that are furthermore {\it decorated by} projection operators.
In the HPO approach itself, time is continuous and there is a~corresponding continuum of projection operators.
Quantum histories commutators $\boldsymbol{[}~, ~\boldsymbol{]}_{\boldsymbol{H}}$ now
replace classical histories brackets, and the f\/inal set of f\/irst-class histories constraints are promoted to quantum
operators $\widehat{{\cal C}}^{\lambda}_{\text{F}}$.
Quantum histories Dirac beables are then further quantum histories operators
$\widehat{D}_{\lambda}^{\text{\sffamily{B}}}$ such that
\begin{gather*}
\boldsymbol{\big[} \widehat{D}_{\text{H}}^{\lambda} , \widehat{\cal C}_{\text{\sffamily{F}}}^{\lambda}\boldsymbol{\big]}_{\boldsymbol{H}}\Psi = 0.   %\label{QHDirObs}
\end{gather*}
On the other hand, for theories with a~subalgebraic structure of quantum linear histories constraints $\widehat{\cal
LIN}_{\text{\sffamily{L}}}^{\lambda}$, quantum histories Kucha\v{r} beables are further quantum histories
operators $\widehat{K}^{\lambda}_{\text{\sffamily{B}}}$ such that
\begin{gather*}
\boldsymbol{\big[} \widehat{K}_{\text{H}}^{\lambda},   \widehat{\cal LIN}_{\text{\sffamily{L}}}^{\lambda}\boldsymbol{\big]}_{\boldsymbol{H}} \Psi = 0.
%\label{QHKObs}
\end{gather*}

\subsection{Notions of brackets with more slots}%\label{Slot}

Suppose instead that the pair $Q^{\text{\sffamily{A}}}$, $P_{\text{\sffamily{A}}}$ of
conventional canonical theory are replaced with $p \geq 3$ dif\/ferent types of entity.
Such a~theory then possesses $p - 1$ scalar entities in place of the standard Hamiltonian.
Such a~theory is additionally equipped with a~generalization of the usual 2-slot notion of brackets to a~bracket
with~$p$ slots.
Even 3-slot theories are relatively novel in concrete physical examples, so I venture no further.

For theories with brackets with three slots, $|\boldsymbol{[}~,~,  \boldsymbol{]}|$, it is these that are to be used to determine which constraints are f\/irst- and second-class, and then also
to form the constraint algebra.
Then additionally the beables notion forming zero brackets with an entity already in the theory (hitherto the
constraints) begins to take multiple forms.
Which of
\begin{gather}
|\boldsymbol{[} {\cal C}_{\text{\sffamily{C}}} ,  {\cal C}_{\text{\sffamily{C}}^{\prime}} ,  B_{\text{\sffamily{B}}}\boldsymbol{]}| \approx 0,
\label{B-1}
\\
|\boldsymbol{[} {\cal C}_{\text{\sffamily{C}}} ,  B_{\text{\sffamily{B}}} ,
  B_{\text{\sffamily{B}}^{\prime}}\boldsymbol{]}| \approx 0
\label{B-2}
\end{gather}
(or both) do we use?   I point out that it is notion~\eqref{B-1} that better parallels the standard notion of conserved
quantity~$C$, which generalizes here to $\textrm{d} C/\textrm{d} t = |\boldsymbol{[} C ,  \text{\sffamily H}
,  \text{\sffamily G} \boldsymbol{]}| \approx 0$ on account of there being two Hamiltonians, $\text{\sffamily
H}$ and $\text{\sffamily G}$.
Perhaps also there are further meaningful notions of {\it bibeables}.
I.e.\
entities $Bi_{\text{\sffamily{V}}}$ such that
\begin{gather*}
|\boldsymbol{[} {\cal C}_{\text{\sffamily{C}}} ,  B_{\text{\sffamily{B}}} ,
  Bi_{\text{\sffamily{V}}} \boldsymbol{]}| \approx 0
\end{gather*}
for whichever of the preceding notions of beables.
This notion is an analogous concept and nomenclature to how a~curve in 3-$d$ has a~binormal in addition to its normal
and its tangent familiar from curves in 2-$d$.
The above concepts about beables for 3-slot brackets indeed suitably extend to~$p$-slot brackets.
This subsection covers objects of which the next two subsections provide important specif\/ic examples from quantum gravity
programs.

\subsection{Associator and Nambu brackets, e.g.\ for M2-branes}%\label{M2}

The {\it associator} bracket is denoted by $\langle ~,~,~\rangle $
and has the algebraic form
\begin{gather*}
\langle  A,  B ,  C \rangle  = A \cdot \{B \cdot C \} - \{A \cdot B \} \cdot C
\end{gather*}
for corresponding notion of plain product operation $\cdot$.
I.e.\
just as the commutator quantif\/ies non-commutativity ($\boldsymbol{[}A,  B\boldsymbol{]} = A \cdot B - B \cdot
A)$, the associator quantif\/ies non-associativity.
The algebra thus formed is indeed nonassociative rather than just noncommutative.

Next, the {\it Nambu bracket}~\cite{Nambu, Takh} is the completely antisymmetrized associator:
\begin{gather*}
\boldsymbol{[}~,~,~\boldsymbol{]} = \langle~,
~,~\rangle  + \text{perms}.
%\label{75}
\end{gather*}
As an important application, the {\it Bagger--Lambert--Gustavsson action}~\cite{M2a, M2b, M2c, M2d} is built out of this
bracket.
This (Example~10) is an action for {\it multiple $N = 8$ M2-branes}\footnote{Here $N =8$ is the amount of supersymmetry
and M2 denotes a~spatially 2-$d$ M-Theory brane.}; this is a~leading candidate for understanding the microphysics of M-Theory.

As a~simpler example, the octonians are a~toy model of the associator bracket.
Here $\boldsymbol{[}u,  v,  w\boldsymbol{]} = \frac{1}{3}\{\{u \times
v\} \times w + \text{cycles}\}$ renders this bracket in terms of
standard mathematics (vector products).
Dimension $>$ 3 is required in order for this not to vanish.
Were it to vanish, the condition for it doing so is that the Jacobi identity of the subsequently in general
noncommutative (but associative) algebra holds.

For theories based on the Nambu brackets, the constraint algebra is of the form
\begin{gather*}
\boldsymbol{[} {\cal C}_{\text{\sffamily{F}}} , {\cal C}_{\text{\sffamily{F}}^{\prime}} ,  {\cal C}_{\text{\sffamily{F}}^{\prime\prime}} \boldsymbol{]} =   f_{\text{\sffamily{F}}\text{\sffamily{F}}^{\prime}\text{\sffamily{F}}^{\prime\prime}}{}^{\text{\sffamily{F}}^{\prime\prime\prime}}{\cal C}_{\text{\sffamily{F}}^{\prime\prime\prime}}.
\end{gather*}
The constant~$f$'s here are manifestly antisymmetric in the f\/irst three indices (and in fact, upon lowering the other
index, totally antisymmetric).
These generalize the standard Lie algebra's structure constants.
Indeed one reason for using Nambu brackets rather than associators themselves is that this generalization of Lie algebra
ensues.
The analogue of Jacobi identity for these brackets as regards which notions of (bi)beables close algebraically is the
{\it Filippov identity}~\cite{Filippov}
\begin{gather*}
\boldsymbol{[} A,  B ,  \boldsymbol{[} C ,  D ,  E \boldsymbol{]}
\boldsymbol{]} = \boldsymbol{[} \boldsymbol{[} A,  B ,  C \boldsymbol{]} ,  D ,
E \boldsymbol{]} + \boldsymbol{[} C ,  \boldsymbol{[} A,  B ,  D \boldsymbol{]} ,
 E \boldsymbol{]} + \boldsymbol{[} C ,  D , \boldsymbol{[} A,  B ,  E   \boldsymbol{]}  \boldsymbol{]}.
\end{gather*}
There is then a~direct parallel of the Jacobi identity working~\eqref{BBC}, by which the Filippov identity and
subsequent use of each of~\eqref{B-1}, \eqref{B-2} or the corresponding notions of bibeable establishes that each of
these notions algebraically closes.
\eqref{B-1}~are then quantities which associate with pairs of constraints, whereas~\eqref{B-2} are presently quantities
pairs of which associate with constraints.
Bibeables are then quantities that associate with whichever f\/irst-class constraint and whichever beable.

\subsection{Master constraint program's double bracket}%\label{Master}

The master constraint program arose in one variant of loop quantum gravity~\cite{Thiemannbook}.
In this approach one has just the usual $Q^{\text{\sffamily{A}}}$ and $P_{\text{\sffamily{A}}}$
and thus just the usual underlying 2-slot brackets notion.
However, one trades one's constraints for a~single {\it master constraint}.
This involves packaging all the constraints (whether or not f\/irst-class~\cite{Thiemannbook}) into a~single object known
as the {\it master constraint}
\begin{gather*}
\mathbb{M}:= \text{$\frac{1}{2}$}\sum_{\text{\sffamily{C}}, \text{\sffamily{C}}^{\prime}} {\cal C}_{\text{\sffamily{C}}}K^{\text{\sffamily{C}}\text{\sffamily{C}}^{\prime}}{\cal C}_{\text{\sffamily{C}}^{\prime}}.
\end{gather*}
Here $K^{\text{\sffamily{C}}\text{\sffamily{C}}^{\prime}}$ is a~positive operator on the space
of square-summable sequences over the index set~$\text{\sffamily C}$.
Then the constraint surface for $\mathbb{M}$ coincides with that for ${\cal C}_{\text{\sffamily{C}}} = 0$.
The master constraint then forms a~trivial constraint algebra with itself.

Moreover, the usual notion of observables/beables is here replaced by the double 2-slot bracket condition
\begin{gather}
\boldsymbol{\{}\text{B}_{\text{\sffamily{B}}} ,  \boldsymbol{\{}B_{\text{\sffamily{B}}^{\prime}}, \mathbb{M}\boldsymbol{\}}\boldsymbol{\}} \approx 0   \label{MCC}
\end{gather}
for {\it master constraint observables/beables}.
By the general identity
\begin{gather*}
\boldsymbol{\{} F ,  \boldsymbol{\{} F ,  \mathbb{M}\boldsymbol{\}}\,\boldsymbol{\}}_{\mathbb{M} = 0} =
\sum_{\text{\sffamily{C}}, \text{\sffamily{C}}^{\prime}}\boldsymbol{ \{}F,  {\cal C}_{\text{\sffamily{C}}} \boldsymbol{\}}_{\mathbb{M} =
0}K^{\text{\sffamily{A}}\text{\sffamily{C}}^{\prime}} \boldsymbol{\{}F,  {\cal
C}_{\text{\sffamily{C}}^{\prime}}\boldsymbol{\}}_{\mathbb{M} = 0},
\end{gather*}
the single master equation~\eqref{MCC} is equivalent to the inf\/inity of equations
$\boldsymbol{\{}B_{\text{\sffamily{B}}} ,  {\cal C}_{\text{\sffamily{C}}} \boldsymbol{\}}_{\mathbb{M} \approx 0} \approx 0$.
Thus the master equation precisely picks out weak   (Dirac)   observables/beables.

This notion indeed also produces a~closed algebraic structure: $\boldsymbol{\{}B_{\text{\sffamily{B}}}
,  B_{\text{\sffamily{B}}^{\prime}} \boldsymbol{\}}$ obeys~\eqref{MCC} if the~$B_{\text{\sffamily{B}}}$ do.
This is by use within the subsequent object built from three Poisson brackets
\begin{gather*}
\boldsymbol{\{} \boldsymbol{\{} B_{\text{\sffamily{B}}} ,  B_{\text{\sffamily{B}}^{\prime}} \boldsymbol{\}} , \boldsymbol{\{} B_{\text{\sffamily{B}}^{\prime\prime}} ,   \mathbb{M} \boldsymbol{\}}\boldsymbol{\}},
\end{gather*}
of f\/irstly the standard Jacobi identity followed by two uses of~\eqref{MCC}.

The idea in this program is then to quantize $\mathbb{M}$ itself.

\section{Conclusion: a~collection of frontiers}%\label{Conclusion}

Types of constraint, types of gauge theory, types of zero, types of bracket have been discussed, and underlie
a~large number of variants on the def\/inition of observables or beables in a~wide range of (models of) gravitational theories.
This remains an open f\/ield.
For instance, the following three programs have largely not yet been considered at the quantum level.

1.~The consequences of Pons et al. classical level work~\cite{PSS09, PSS10}.

2.~Dittrich's power series~\cite{Dittrich1, Dittrich2}.

3.~Concrete examples of Histories Theory approaches with nontrivial dif\/feomorphisms~\cite{Kouletsis,
Savvidou04b}.

4.~On the other hand, Halliwell's approach~\cite{H03, H09, H10} has not yet been applied to perturbative or
exact midisuperspace models; the more advanced form of this~\cite{H09, H10} has not even been applied to any examples
with additional linear gauge constraints.

5.~Halliwell and Marolf's~\cite{Marolf09} approaches remain to be compared in detail with each other.
Does one of these confer greater advantages, do the two approaches produce compatible results?

6.~Footnote~\ref{foo-log} considers various further areas ripe for review that are not covered in the present
review or recent previous ones~\cite{PSS10, Tambornino}.
One would hope then that the population of such recent reviews would grow by a~further one or two due to someone else's
further treatment of asymptotic/boundary/holographic observables/beables in gravitational theories.

7.~As outlined in this review, the beables aspect of supergravity is still in its infancy.

8.~This review only touches on one of many possible applications of observables/beables to M-theory: following
on from the use of Nambu brackets in the Bagger--Lambert--Gustavsson action.

Resolving the quantum problem of beables (and in a~manner consistent with the rest of the problem of time
facets of Appendix~\ref{PoT}) remains a~major and overarching open problem.

\appendix

\section{Facets of the problem of time}\label{PoT}

The problem of time~\cite{APOT2, I93, Kuchar92} has 9 facets stemming together from the mismatch in time concepts
between GR and QM.
See~\cite{BI} for a~more detailed account of these and how they are each underlied by an aspect of background
independence.

{\bf Facet 1.} {\it Frozen formalism problem}.
The Schr\"{o}dinger picture manifestation is that GR's quantum wave equation~-- the Wheeler--DeWitt equation
$\widehat{\cal H}\Psi = 0$, is a~stationary, i.e.~timeless wave equation, like $\widehat{H}\Psi = E\Psi$ as opposed to
$\widehat{H}\Psi = i\hbar\partial\Psi/\partial t$ for some notion of time~$t$.
This question is determined by the purely quadratic nature of GR's Hamiltonian constraint ${\cal H}$, and in turn by the
Leibnizian `there is no time for the universe as a~whole' {\it temporal relationalism principle}.
Strategies for resolving this facet include the following.

O.~Perhaps a~fundamental hidden/internal time that can be discovered within classical GR by canonical reformulation.

A.~Perhaps GR has emergent rather than fundamental time.
E.g.\ in the semiclassical regime, slow, heavy modes provide an approximate time by which the other fast, light modes evolve.

B.~Instead one could see how much of physics can be done timelessly, e.g.\ via addressing solely questions of being, rather than becoming.
For example, {\it records theory}~\cite{ARec1, FileR, ARec2} concerns whether a~single instant contains
pattern/correlation information, and whether a~semblance of dynamics or history arises from this.

C.~Perhaps histories are primary instead: {\it histories theory}.
At the quantum level, this involves consistent sets of histories, f\/ine- and coarse-graining operations and decoherence
functionals for comparing pairs of histories.
Some versions~\cite{IL} have furthermore a~classical precursor involving the histories brackets as outlined in
Section~\ref{HB}.

One can furthermore combine schemes A-C~\cite{AHall, H03, H09} since   i)~histories contain records,
ii)~histories decohereing (self-measuring) gives semiclassicality,
iii)~the elusive question of which degrees of freedom decohere which others is addressed via where info is actually
stored, i.e.~where the records are.
Such a~combination underlies Halliwell's approach to the problem of time~\cite{H03, H09} of Sections~\ref{Dir-Ex},~\ref{Hall-1} and \ref{Hall-2}.

D.~Consider QM in the Heisenberg picture instead (cf.\
Section~\ref{Hall-1}), whether as a~more lucid choice or as exclusively the only choice for quantum gravity~\cite{Rovellibook}.

{\bf Facet 2} -- {\it configurational relationalism}~-- concerns dynamics with a~physically irrelevant group of
transformations $\mathfrak{g}$ acting on conf\/iguration space~$\mathfrak{q}$.
This gives rise to constraints that are linear in the momenta.
Conf\/igurational relationalism is resolved for relational particle mechanics by Barbour's {\it best matching}: solving
the Lagrangian form of the linear constraints for the $\mathfrak{g}$-auxiliary variables themselves.
(This is an example of Section~\ref{K-B} formal strategy~1.)   However this remains unresolved as Wheeler's well-known
{\it thin Sandwich problem}~\cite{BSW, Sandwich} occurs in the case of GR, for which the linear constraint is ${\cal
M}_i$, $\mathfrak{g}$ is Dif\/f$(\Sigma)$ and the $\mathfrak{g}$-auxiliary variables form
the shift vector~$N^i$.

{\bf Facet 3} is the {\it constraint closure problem}, i.e.~whether the brackets of the constraints~${\cal H}$ and~${\cal M}_i$ do not produce further conditions as per~\eqref{MM}--\eqref{HH}.

{\bf Facet 4} is the problem of beables/observables as exposited in the Introduction.

{\bf Facet 5} is {\it spacetime relationalism}.
Here a~physically-irrelevant group of motions acts on spacetime $\mathfrak{m}$; this is usually
Dif\/f($\mathfrak{m}$).
This problem picks up additional nontrivialities as regards representations and the measures in path-integral and
histories formulations at the quantum level.

{\bf Facet 6} is what happens at the quantum level to GR's independence of foliation of spacetime by spaces.
This corresponds to the theoretical scheme being able to encode arbitrarily moving families of observers.

{\bf Facet 7} concerns how classical spacetime is to be reconstructed~\cite{AM13} from space and/or discrete notions.

Note that the Dirac algebroid for ${\cal H}$ and ${\cal M}_i$ resolves all three of~3),~6),~7) at the
classical level.
However even the semiclassical counterpart of these resolutions remains unknown.

{\bf Facet 8} is a~collection~\cite{GPoT} of global complications with each of timefunctions (that could refer to
globality in time itself or in how time is def\/ined over space), or the other facets above.
(E.g.\ best matching, foliations and spacetime reconstruction are in general only local constructs and with strategies for
resolving facets (e.g.\ only locally def\/ined hidden and emergent times.)

{\bf Facet 9} is the {\it multiple choice problem}: that classically canonically equivalent formulations are in
general unitarily inequivalent at the quantum level~\cite{Gotay}.

\subsection*{Acknowledgements}

E.A.~thanks close people, Julian Barbour, Jeremy Butterf\/ield, Harvey Brown, Sean Gryb, Jonathan Halliwell, Philipp
H\"ohn, Chris Isham, Flavio Mercati, Brian Pitts, Josep Maria Pons, Oliver Pooley, Donald Salisbury, Dimitri Vey, Hans
Westman, Michael Wright and the anonymous referees for discussions, Jeremy Butterf\/ield, John Barrow, Marc
Lachi\'eze--Rey, Malcolm MacCallum, Don Page, Reza Tavakol, Juan Valiente-Kroon and Paulo Vargas-Moniz
for help with my career, and DAMTP Cambridge, Perimeter Institute Waterloo and the University of New Brunswick
Fredericton for hospitality at various points during the making of this %Invited
review.
This work started within my grant from the Foundational Questions Institute (FQXi) Fund, a~donor-advised fund of the
Silicon Valley Community Foundation on the basis of proposal FQXi-RFP3-1101 to the FQXi, administered via Theiss
Research and the CNRS and held at APC Universit\'{e} Paris Diderot.

\pdfbookmark[1]{References}{ref}
\LastPageEnding


\begin{thebibliography}{99}
\footnotesize\itemsep=-0.5pt

\bibitem{Ana01}
Anastopoulos C., Continuous-time histories: observables, probabilities, phase
  space structure and the classical limit, \href{http://dx.doi.org/10.1063/1.1383975}{\textit{J.~Math. Phys.}} \textbf{42}
  (2001), 3225--3259, \href{http://arxiv.org/abs/quant-ph/0008052}{quant-ph/0008052}.

\bibitem{ARec1}
Anderson E., Records theory, \href{http://dx.doi.org/10.1142/S0218271809014686}{\textit{Internat.~J. Modern Phys.~D}} \textbf{18}
  (2009), 635--667, \href{http://arxiv.org/abs/0709.1892}{arXiv:0709.1892}.

\bibitem{AHall}
Anderson E., Approaching the problem of time with a combined
  semiclassical-records-histories scheme, \href{http://dx.doi.org/10.1088/0264-9381/29/23/235015}{\textit{Classical Quantum Gravity}}
  \textbf{29} (2012), 235015, 37~pages, \href{http://arxiv.org/abs/1204.2868}{arXiv:1204.2868}.

\bibitem{APOT2}
Anderson E., Problem of time in quantum gravity, \href{http://dx.doi.org/10.1002/andp.201200147}{\textit{Ann. Phys.}}
  \textbf{524} (2012), 757--786, \href{http://arxiv.org/abs/1206.2403}{arXiv:1206.2403}.

\bibitem{APOT}
Anderson E., The problem of time in quantum gravity, in Classical and Quantum
  Gravity: Theory, Analysis and Applications, Editor V.R.~Frignanni, Nova, New
  York, 2012, 213--256, \href{http://arxiv.org/abs/1009.2157}{arXiv:1009.2157}.

\bibitem{QuadI}
Anderson E., Relational quadrilateralland. {I}.~{T}he classical theory,
  \href{http://dx.doi.org/10.1142/S021827181450014X}{\textit{Internat.~J. Modern Phys.~D}} \textbf{23} (2014), 1450014, 75~pages,
  \href{http://arxiv.org/abs/1202.4186}{arXiv:1202.4186}.

\bibitem{FileR}
Anderson E., The problem of time and quantum cosmology in the relational
  particle mechanics arena, \href{http://arxiv.org/abs/1111.1472}{arXiv:1111.1472}.

\bibitem{ARel2}
Anderson E., Machian time is to be abstracted from what change?,
  \href{http://arxiv.org/abs/1209.1266}{arXiv:1209.1266}.

\bibitem{ARec2}
Anderson E., Kendall's shape statistics as a classical realization of
  Barbour-type timeless records theory approach to quantum gravity,
  \href{http://arxiv.org/abs/1307.1923}{arXiv:1307.1923}.

\bibitem{BI}
Anderson E., Background independence, \href{http://arxiv.org/abs/1310.1524}{arXiv:1310.1524}.

\bibitem{ABrackets}
Anderson E., Problem of time in slightly inhomogeneous cosmology,
  \href{http://arxiv.org/abs/1403.7583}{arXiv:1403.7583}.

\bibitem{GPoT}
Anderson E., Global problems of time in quantum gravity, in preparation.

\bibitem{AM13}
Anderson E., Mercati F., Classical Machian resolution of the spacetime
  reconstruction problem, \href{http://arxiv.org/abs/1311.6541}{arXiv:1311.6541}.

\bibitem{+Torre2}
Anderson I.M., Torre C.G., Classif\/ication of local generalized symmetries for
  the vacuum {E}instein equations, \href{http://dx.doi.org/10.1007/BF02099248}{\textit{Comm. Math. Phys.}} \textbf{176}
  (1996), 479--539, \href{http://arxiv.org/abs/gr-qc/9404030}{gr-qc/9404030}.

\bibitem{A67}
Anderson J.L., Principles of relativity physics, Academic Press, New York,
  1967.

\bibitem{AB51}
Anderson J.L., Bergmann P.G., Constraints in covariant f\/ield theories,
  \href{http://dx.doi.org/10.1103/PhysRev.83.1018}{\textit{Phys. Rev.}} \textbf{83} (1951), 1018--1025.

\bibitem{Arnol'd}
Arnold V.I., Mathematical methods of classical mechanics, \textit{Graduate
  Texts in Mathematics}, Vol.~60, Springer-Verlag, New York~-- Heidelberg,
  1978.

\bibitem{A91}
Ashtekar A., Lectures on nonperturbative canonical gravity, \href{http://dx.doi.org/10.1142/1321}{\textit{Advanced
  Series in Astrophysics and Cosmology}}, Vol.~6, World Sci. Publ., River Edge,
  NJ, 1991.

\bibitem{ATU93}
Ashtekar A., Tate R., Uggla C., Minisuperspaces: observables and quantization,
  \href{http://dx.doi.org/10.1142/S0218271893000039}{\textit{Internat.~J. Modern Phys.~D}} \textbf{2} (1993), 15--50,
  \href{http://arxiv.org/abs/gr-qc/9302027}{gr-qc/9302027}.

\bibitem{M2a}
Bagger J., Lambert N., Modeling multiple {M}2-branes, \href{http://dx.doi.org/10.1103/PhysRevD.75.045020}{\textit{Phys. Rev.~D}}
  \textbf{75} (2007), 045020, 7~pages, \href{http://arxiv.org/abs/hep-th/0611108}{hep-th/0611108}.

\bibitem{M2b}
Bagger J., Lambert N., Comments on multiple {M}2-branes, \href{http://dx.doi.org/10.1088/1126-6708/2008/02/105}{\textit{J.~High Energy
  Phys.}} \textbf{2008} (2008), no.~2, 105, 15~pages, \href{http://arxiv.org/abs/0712.3738}{arXiv:0712.3738}.

\bibitem{M2c}
Bagger J., Lambert N., Three-algebras and {${\mathcal N}=6$} {C}hern--{S}imons
  gauge theories, \href{http://dx.doi.org/10.1103/PhysRevD.79.025002}{\textit{Phys. Rev.~D}} \textbf{79} (2009), 025002, 8~pages,
  \href{http://arxiv.org/abs/0807.0163}{arXiv:0807.0163}.

\bibitem{BSW}
Baierlein R.F., Sharp D.H., Wheeler J.A., Three-dimensional geometry as carrier
  of information about time, \href{http://dx.doi.org/10.1103/PhysRev.126.1864}{\textit{Phys. Rev.}} \textbf{126} (1962),
  1864--1866.

\bibitem{EOT}
Barbour J., The end of time, Oxford University Press, Oxford, 2000.

\bibitem{BFOA}
Barbour J., Foster B.Z., {\'O}~Murchadha N., Relativity without relativity,
  \href{http://dx.doi.org/10.1088/0264-9381/19/12/308}{\textit{Classical Quantum Gravity}} \textbf{19} (2002), 3217--3248,
  \href{http://arxiv.org/abs/gr-qc/0012089}{gr-qc/0012089}.

\bibitem{B94I}
Barbour J.B., The timelessness of quantum gravity. {I}.~{T}he evidence from the
  classical theory, \href{http://dx.doi.org/10.1088/0264-9381/11/12/005}{\textit{Classical Quantum Gravity}} \textbf{11} (1994),
  2853--2873.

\bibitem{B94II}
Barbour J.B., The timelessness of quantum gravity. {II}.~{T}he appearance of
  dynamics in static conf\/igurations, \href{http://dx.doi.org/10.1088/0264-9381/11/12/006}{\textit{Classical Quantum Gravity}}
  \textbf{11} (1994), 2875--2897.

\bibitem{BB82}
Barbour J.B., Bertotti B., Mach's principle and the structure of dynamical
  theories, \href{http://dx.doi.org/10.1098/rspa.1982.0102}{\textit{Proc. Roy. Soc. London Ser.~A}} \textbf{382} (1982),
  295--306.

\bibitem{BF08}
Barbour J.B., Foster B.Z., Constraints and gauge transformations: Dirac's
  theorem is not always valid, \href{http://arxiv.org/abs/0808.1223}{arXiv:0808.1223}.

\bibitem{Bardeen}
Bardeen J.M., Gauge-invariant cosmological perturbations, \href{http://dx.doi.org/10.1103/PhysRevD.22.1882}{\textit{Phys. Rev.~D}}
  \textbf{22} (1980), 1882--1905.

\bibitem{Bohm8}
Barrett J., Wigner's friend and Bell's f\/ield beables, in Vision of Oneness,
  Editors I.~Licata, A.~Sakaji, Aracne, Rome, 2011, 63--82.

\bibitem{Sandwich}
Bartnik R., Fodor G., On the restricted validity of the thin sandwich
  conjecture, \href{http://dx.doi.org/10.1103/PhysRevD.48.3596}{\textit{Phys. Rev.~D}} \textbf{48} (1993), 3596--3599.

\bibitem{BT91}
Batalin I.A., Tyutin I.V., Existence theorem for the ef\/fective gauge algebra in
  the generalized canonical formalism with abelian conversion of second-class
  constraints, \href{http://dx.doi.org/10.1142/S0217751X91001581}{\textit{Internat.~J. Modern Phys. A}} \textbf{6} (1991),
  3255--3282.

\bibitem{Bell}
Bell J.S., Quantum mechanics for cosmologists, in Second Oxford Symposium on
  Quantum Gravity, Editors C.J.~Isham, R.~Penrose, D.W.~Sciama, Clarendon,
  Oxford, 1981, 611--637.

\bibitem{Bell87}
Bell J.S., Beables for quantum f\/ield theory, in Quantum Implications, Editors
  B.J.~Hiley, D.~Peat, Routledge \& Kegan Paul, London, 1987, 227--234.

\bibitem{Bell75}
Bell J.S., The theory of local beables, in Speakable and Unspeakable in Quantum
  Mechanics, Cambridge University Press, Cambridge, 1987, 52--62.

\bibitem{Ear01}
Belot G., Earman J., Pre-socratic quantum gravity, in Physics Meets Philosophy
  at the Planck Scale, Editors C.~Callendar, N.~Huggett, Cambridge University
  Press, Cambridge, 2001, 213--255.

\bibitem{Bergmann61}
Bergmann P.G., ``{G}auge-invariant'' variables in general relativity,
  \href{http://dx.doi.org/10.1103/PhysRev.124.274}{\textit{Phys. Rev.}} \textbf{124} (1961), 274--278.

\bibitem{BK72}
Bergmann P.G., Komar A., The coordinate group symmetries of general relativity,
  \href{http://dx.doi.org/10.1007/BF00671650}{\textit{Internat.~J. Theoret. Phys.}} \textbf{5} (1972), 15--28.

\bibitem{Diffs2}
Berman D.S., Cederwall M., Kleinschmidt A., Thompson D.C., The gauge structure
  of generalised dif\/feomorphisms, \href{http://dx.doi.org/10.1007/JHEP01(2013)064}{\textit{J.~High Energy Phys.}} \textbf{2013}
  (2013), no.~1, 064, 22~pages, \href{http://arxiv.org/abs/1208.5884}{arXiv:1208.5884}.

\bibitem{Bleecker}
Bleecker D., Gauge theory and variational principles, Dover, New York, 2001.

\bibitem{BojoBook}
Bojowald M., Canonical gravity and applications: cosmology, black holes, and
  quantum gravity, Cambridge University Press, Cambridge, 2011.

\bibitem{Bojo1}
Bojowald M., H{\"o}hn P.A., Tsobanjan A., An ef\/fective approach to the problem
  of time, \href{http://dx.doi.org/10.1088/0264-9381/28/3/035006}{\textit{Classical Quantum Gravity}} \textbf{28} (2011), 035006,
  18~pages, \href{http://arxiv.org/abs/1009.5953}{arXiv:1009.5953}.

\bibitem{Bojo2}
Bojowald M., H{\"o}hn P.A., Tsobanjan A., Ef\/fective approach to the problem of
  time: general features and examples, \href{http://dx.doi.org/10.1103/PhysRevD.83.125023}{\textit{Phys. Rev.~D}} \textbf{83}
  (2011), 125023, 38~pages, \href{http://arxiv.org/abs/1011.3040}{arXiv:1011.3040}.

\bibitem{Bohm2}
Bub J., Why not take all observables as beables?, in Fundamental Problems in
  Quantum Theory ({B}altimore, {MD}, 1994), \href{http://dx.doi.org/10.1111/j.1749-6632.1995.tb39018.x}{\textit{Ann. New York Acad. Sci.}},
  Vol.~755, New York Acad. Sci., New York, 1995, 761--767.

\bibitem{Algebroid1}
Cannas~da Silva A., Weinstein A., Geometric models for noncommutative algebras,
  \textit{Berkeley Mathematics Lecture Notes}, Vol.~10, Amer. Math. Soc.,
  Providence, RI, 1999.

\bibitem{Carlip90}
Carlip S., Observables, gauge invariance, and time in {$(2+1)$}-dimensional
  quantum gravity, \href{http://dx.doi.org/10.1103/PhysRevD.42.2647}{\textit{Phys. Rev.~D}} \textbf{42} (1990), 2647--2654.

\bibitem{Carlip91b}
Carlip S., Measuring the metric in {$(2+1)$}-dimensional quantum gravity,
  \href{http://dx.doi.org/10.1088/0264-9381/8/1/007}{\textit{Classical Quantum Gravity}} \textbf{8} (1991), 5--17.

\bibitem{Carlip01}
Carlip S., Quantum gravity: a progress report, \href{http://dx.doi.org/10.1088/0034-4885/64/8/301}{\textit{Rep. Progr. Phys.}}
  \textbf{64} (2001), 885--942, \href{http://arxiv.org/abs/gr-qc/0108040}{gr-qc/0108040}.

\bibitem{Casalbuoni}
Casalbuoni R., On the quantization of systems with anticommuting variables,
  \href{http://dx.doi.org/10.1007/BF02748689}{\textit{Nuovo Cimento~A}} \textbf{33} (1976), 115--125.

\bibitem{Bohm4}
Clifton R., Beables in algebraic quantum theory, in From Physics to Philosophy
  ({C}ambridge, 1997), Editors M.~Redhead, J.~Butterf\/ield, C.~Pagonis,
  Cambridge University Press, Cambridge, 1999, 12--43.

\bibitem{DEath}
D'Eath P.D., Supersymmetric quantum cosmology, \href{http://dx.doi.org/10.1017/CBO9780511524424}{\textit{Cambridge Monographs on
  Mathematical Physics}}, Cambridge University Press, Cambridge, 1996.

\bibitem{Des}
Deser S., Kay J.H., Stelle K.S., Hamiltonian formulation of supergravity,
  \href{http://dx.doi.org/10.1103/PhysRevD.16.2448}{\textit{Phys. Rev.~D}} \textbf{16} (1977), 2448--2455.

\bibitem{DeWitt62}
DeWitt B.S., The quantization of geometry, in Gravitation: {A}n Introduction to
  Current Research, Editor L.~Witten, Wiley, New York, 1962, 266--381.

\bibitem{DiracObs}
Dirac P.A.M., Forms of relativistic dynamics, \href{http://dx.doi.org/10.1103/RevModPhys.21.392}{\textit{Rev. Modern Phys.}}
  \textbf{21} (1949), 392--399.

\bibitem{Dirac}
Dirac P.A.M., Lectures on quantum mechanics, \textit{Belfer Graduate School of
  Science Monographs Series}, Vol.~2, Belfer Graduate School of Science, New
  York, 1964.

\bibitem{Dittrich1}
Dittrich B., Partial and complete observables for canonical general relativity,
  \href{http://dx.doi.org/10.1088/0264-9381/23/22/006}{\textit{Classical Quantum Gravity}} \textbf{23} (2006), 6155--6184,
  \href{http://arxiv.org/abs/gr-qc/0507106}{gr-qc/0507106}.

\bibitem{Dittrich2}
Dittrich B., Partial and complete observables for {H}amiltonian constrained
  systems, \href{http://dx.doi.org/10.1007/s10714-007-0495-2}{\textit{Gen. Relativity Gravitation}} \textbf{39} (2007), 1891--1927,
  \href{http://arxiv.org/abs/gr-qc/0411013}{gr-qc/0411013}.

\bibitem{DittTamb2}
Dittrich B., Tambornino J., Gauge-invariant perturbations around
  symmetry-reduced sectors of general relativity: applications to cosmology,
  \href{http://dx.doi.org/10.1088/0264-9381/24/18/001}{\textit{Classical Quantum Gravity}} \textbf{24} (2007), 4543--4585,
  \href{http://arxiv.org/abs/gr-qc/0702093}{gr-qc/0702093}.

\bibitem{DittTamb1}
Dittrich B., Tambornino J., A perturbative approach to {D}irac observables and
  their spacetime algebra, \href{http://dx.doi.org/10.1088/0264-9381/24/4/001}{\textit{Classical Quantum Gravity}} \textbf{24}
  (2007), 757--783, \href{http://arxiv.org/abs/gr-qc/0610060}{gr-qc/0610060}.

\bibitem{ID}
D\"oring A., Isham C., ``{W}hat is a thing?'': topos theory in the foundations
  of physics, in New Structures for Physics, \href{http://dx.doi.org/10.1007/978-3-642-12821-9_13}{\textit{Lecture Notes in Phys.}},
  Vol.~813, Editor R.~Coecke, Springer, Heidelberg, 2011, 753--937,
  \href{http://arxiv.org/abs/0803.0417}{arXiv:0803.0417}.

\bibitem{Ear02}
Earman J., Gauge matters, \href{http://dx.doi.org/10.1086/341847}{\textit{Philos. Sci.}} \textbf{69} (2002), S209--S220.

\bibitem{Filippov}
Filippov V.T., $n$-Lie algebras, \href{http://dx.doi.org/10.1007/BF00969110}{\textit{Sib. Math.~J.}} \textbf{26} (1985),
  879--891.

\bibitem{FV}
Fradkin E.S., Vasiliev M.A., Hamiltonian formalism, quantization and {$S$}
  matrix for supergravity, \href{http://dx.doi.org/10.1016/0370-2693(77)90065-X}{\textit{Phys. Lett.~B}} \textbf{72} (1977), 70--74.

\bibitem{GPP1}
Gambini R., Porto R.A., Relational time in generally covariant quantum systems:
  four models, \href{http://dx.doi.org/10.1103/PhysRevD.63.105014}{\textit{Phys. Rev.~D}} \textbf{63} (2001), 105014, 15~pages,
  \href{http://arxiv.org/abs/gr-qc/0101057}{gr-qc/0101057}.

\bibitem{GPP2}
Gambini R., Porto R.A., Pullin J., Torterolo S., Conditional probabilities with
  {D}irac observables and the problem of time in quantum gravity, \href{http://dx.doi.org/10.1103/PhysRevD.79.041501}{\textit{Phys.
  Rev.~D}} \textbf{79} (2009), 041501, 5~pages, \href{http://arxiv.org/abs/0809.4235}{arXiv:0809.4235}.

\bibitem{GPbook}
Gambini R., Pullin J., Loops, knots, gauge theories and quantum gravity,
  \href{http://dx.doi.org/10.1017/CBO9780511524431}{\textit{Cambridge Monographs on Ma\-the\-matical Physics}}, Cambridge University Press,
  Cambridge, 1996.

\bibitem{PG05}
Gambini R., Pullin J., Consistent discrete space-time, in 100 Years of
  Relativity. Space-Time Structure: Einstein and Beyond, Editor A.~Ashtekar,
  \href{http://dx.doi.org/10.1142/9789812700988_0015}{World Sci. Publ.}, Hackensack, NJ, 2005, 415--444.

\bibitem{DG}
G\'{e}h\'{e}niau J., Debever R., Les quatorze invariants de courbure de
  l'espace Riemannien a quatre dimensions, \textit{Helv. Phys. Acta Suppl.}
  \textbf{4} (1956), 101--105.

\bibitem{GMH}
Gell-Mann M., Hartle J.B., Classical equations for quantum systems,
  \href{http://dx.doi.org/10.1103/PhysRevD.47.3345}{\textit{Phys. Rev.~D}} \textbf{47} (1993), 3345--3382, \href{http://arxiv.org/abs/gr-qc/9210010}{gr-qc/9210010}.

\bibitem{G-M-H}
Giddings S.B., Marolf D., Hartle J.B., Observables in ef\/fective gravity,
  \href{http://dx.doi.org/10.1103/PhysRevD.74.064018}{\textit{Phys. Rev.~D}} \textbf{74} (2006), 064018, 20~pages,
  \href{http://arxiv.org/abs/hep-th/0512200}{hep-th/0512200}.

\bibitem{Bohm9}
Goldstein S., Norsen T., Tausk D.V., Zanghi N., Bell's theorem,
  \href{http://dx.doi.org/10.4249/scholarpedia.8378}{\textit{Scholarpedia}} \textbf{6} (2011), 8378.

\bibitem{Gotay}
Gotay M.J., Obstructions to quantization, in Mechanics: from Theory to
  Computation (Essays in Honor of Juan-Carlos Sim\'{o}, Editors J.~Marsden,
  S.~Wiggins, Springer, New York, 2000, 171--216, \href{http://arxiv.org/abs/math-ph/9809011}{math-ph/9809011}.

\bibitem{M2d}
Gustavsson A., Algebraic structures on parallel {M}2 branes, \href{http://dx.doi.org/10.1016/j.nuclphysb.2008.11.014}{\textit{Nuclear
  Phys.~B}} \textbf{811} (2009), 66--76, \href{http://arxiv.org/abs/0709.1260}{arXiv:0709.1260}.

\bibitem{+Perennials1}
H{\'a}j{\'{\i}}{\v{c}}ek P., Group quantization of parametrized systems.
  {I}.~{T}ime levels, \href{http://dx.doi.org/10.1063/1.530911}{\textit{J.~Math. Phys.}} \textbf{36} (1995), 4612--4638,
  \href{http://arxiv.org/abs/gr-qc/9412047}{gr-qc/9412047}.

\bibitem{+Perennials2}
H{\'a}j{\'{\i}}{\v{c}}ek P., Isham C.J., Perennials and the group-theoretical
  quantization of a parametrized scalar f\/ield on a curved background,
  \href{http://dx.doi.org/10.1063/1.531579}{\textit{J.~Math. Phys.}} \textbf{37} (1996), 3522--3538,
  \href{http://arxiv.org/abs/gr-qc/9510034}{gr-qc/9510034}.

\bibitem{H03}
Halliwell J., The interpretation of quantum cosmology and the problem of time,
  in The Future of the Theoretical Physics and Cosmology ({C}ambridge, 2002),
  Editors G.W.~Gibbons, E.P.S.~Shellard, S.J.~Rankin, Cambridge University
  Press, Cambridge, 2003, 675--692, \href{http://arxiv.org/abs/gr-qc/0208018}{gr-qc/0208018}.

\bibitem{H99}
Halliwell J.J., Somewhere in the universe: where is the information stored when
  histories decohere?, \href{http://dx.doi.org/10.1103/PhysRevD.60.105031}{\textit{Phys. Rev.~D}} \textbf{60} (1999), 105031,
  17~pages, \href{http://arxiv.org/abs/quant-ph/9902008}{quant-ph/9902008}.

\bibitem{H09}
Halliwell J.J., Probabilities in quantum cosmological models: a~decoherent
  histories analysis using a complex potential, \href{http://dx.doi.org/10.1103/PhysRevD.80.124032}{\textit{Phys. Rev.~D}}
  \textbf{80} (2009), 124032, 21~pages, \href{http://arxiv.org/abs/0909.2597}{arXiv:0909.2597}.

\bibitem{H10}
Halliwell J.J., Decoherent histories analysis of minisuperspace quantum
  cosmology, \href{http://dx.doi.org/10.1088/1742-6596/306/1/012023}{\textit{J.~Phys. Conf. Ser.}} \textbf{306} (2011), 012023, 22~pages, \href{http://arxiv.org/abs/1108.5991}{arXiv:1108.5991}.

\bibitem{HallHaw1}
Halliwell J.J., Hawking S.W., Origin of structure in the {U}niverse,
  \href{http://dx.doi.org/10.1103/PhysRevD.31.1777}{\textit{Phys. Rev.~D}} \textbf{31} (1985), 1777--1791.

\bibitem{HT02}
Halliwell J.J., Thorwart J., Life in an energy eigenstate: decoherent histories
  analysis of a model timeless universe, \href{http://dx.doi.org/10.1103/PhysRevD.65.104009}{\textit{Phys. Rev.~D}} \textbf{65}
  (2002), 104009, 19~pages, \href{http://arxiv.org/abs/gr-qc/0201070}{gr-qc/0201070}.

\bibitem{Bohm3}
Halvorson H., Clifton R., Maximal beable subalgebras of quantum mechanical
  observables, \href{http://dx.doi.org/10.1023/A:1026628407645}{\textit{Internat.~J. Theoret. Phys.}} \textbf{38} (1999),
  2441--2484, \href{http://arxiv.org/abs/quant-ph/9905042}{quant-ph/9905042}.

\bibitem{Hartle95}
Hartle J.B., Spacetime quantum mechanics and the quantum mechanics of
  spacetime, in Gravitation et Quantif\/ications ({L}es {H}ouches, 1992), Editors
  B.~Julia, J.~Zinn-Justin, North-Holland, Amsterdam, 1995, 285--480,
  \href{http://arxiv.org/abs/gr-qc/9304006}{gr-qc/9304006}.

\bibitem{HallHaw5}
Hawking S.W., Laf\/lamme R., Lyons G.W., Origin of time asymmetry, \href{http://dx.doi.org/10.1103/PhysRevD.47.5342}{\textit{Phys.
  Rev.~D}} \textbf{47} (1993), 5342--5356, \href{http://arxiv.org/abs/gr-qc/9301017}{gr-qc/9301017}.

\bibitem{NSI1}
Hawking S.W., Page D.N., Operator ordering and the f\/latness of the universe,
  \href{http://dx.doi.org/10.1016/0550-3213(86)90478-5}{\textit{Nuclear Phys.~B}} \textbf{264} (1986), 185--196.

\bibitem{Multisymplectic}
H{\'e}lein F., Kouneiher J., The notion of observable in the covariant
  {H}amiltonian formalism for the calculus of variations with several
  variables, \href{http://dx.doi.org/10.4310/ATMP.2004.v8.n4.a4}{\textit{Adv. Theor. Math. Phys.}} \textbf{8} (2004), 735--777,
  \href{http://arxiv.org/abs/math-ph/0401047}{math-ph/0401047}.

\bibitem{HT92}
Henneaux M., Teitelboim C., Quantization of gauge systems, Princeton University
  Press, Princeton, NJ, 1992.

\bibitem{Bojo3}
H\"ohn P.A., Ef\/fective relational dynamics, \href{http://dx.doi.org/10.1088/1742-6596/360/1/012014}{\textit{J.~Phys. Conf. Ser.}}
  \textbf{360} (2012), 012014, 4~pages, \href{http://arxiv.org/abs/1110.5631}{arXiv:1110.5631}.

\bibitem{Hoehn}
H\"ohn P.A., From classical to quantum: new canonical tools for the dynamics of
  gravity, Ph.D.\ Thesis, Utrecht University, 2012.

\bibitem{Bojo4}
H\"ohn P.A., Kubalov\'a E., Tsobanjan A., Ef\/fective relational dynamics of a
  nonintegrable cosmological model, \href{http://dx.doi.org/10.1103/PhysRevD.86.065014}{\textit{Phys. Rev.~D}} \textbf{86} (2012),
  065014, 21~pages, \href{http://arxiv.org/abs/1111.5193}{arXiv:1111.5193}.

\bibitem{I84}
Isham C.J., Topological and global aspects of quantum theory, in Relativity,
  Groups and Topology,~{II} ({L}es {H}ouches, 1983), Editors B.~DeWitt,
  R.~Stora, North-Holland, Amsterdam, 1984, 1059--1290.

\bibitem{I93}
Isham C.J., Canonical quantum gravity and the problem of time, in Integrable
  Systems, Quantum Groups, and Quantum Field Theories ({S}alamanca, 1992),
  \textit{NATO Adv. Sci. Inst. Ser. C Math. Phys. Sci.}, Vol.~409, Editors L.A.~Ibort, M.A.~Rodr\'{\i}guez, Kluwer Acad. Publ., Dordrecht, 1993, 157--287,
  \href{http://arxiv.org/abs/gr-qc/9210011}{gr-qc/9210011}.

\bibitem{IBook}
Isham C.J., Lectures on quantum theory. Mathematical and structural
  foundations, \href{http://dx.doi.org/10.1142/p001}{Imperial College Press}, London, 1995.

\bibitem{IK85I}
Isham C.J., Kucha{\v{r}} K.V., Representations of spacetime dif\/feomorphisms.
  {I}.~{C}anonical parametrized f\/ield theories, \href{http://dx.doi.org/10.1016/0003-4916(85)90018-1}{\textit{Ann. Physics}}
  \textbf{164} (1985), 288--315.

\bibitem{IK85II}
Isham C.J., Kucha{\v{r}} K.V., Representations of spacetime dif\/feomorphisms.
  {II}.~{C}anonical geometrodynamics, \href{http://dx.doi.org/10.1016/0003-4916(85)90019-3}{\textit{Ann. Physics}} \textbf{164}
  (1985), 316--333.

\bibitem{IL}
Isham C.J., Linden N., Continuous histories and the history group in
  generalized quantum theory, \href{http://dx.doi.org/10.1063/1.531267}{\textit{J.~Math. Phys.}} \textbf{36} (1995),
  5392--5408, \href{http://arxiv.org/abs/gr-qc/9503063}{gr-qc/9503063}.

\bibitem{JP09}
Jizba P., Pons J.M., Revisiting the gauge principle: enforcing constants of
  motion as constraints, \href{http://dx.doi.org/10.1088/1751-8113/43/20/205202}{\textit{J.~Phys.~A: Math. Theor.}} \textbf{43} (2010),
  205202, 20~pages, \href{http://arxiv.org/abs/0905.3807}{arXiv:0905.3807}.

\bibitem{JZ}
Joos E., Zeh H.D., The emergence of classical properties through interaction
  with the environment, \href{http://dx.doi.org/10.1007/BF01725541}{\textit{Z.~Phys.~B}} \textbf{59} (1985), 223--243.

\bibitem{Giu}
Joos E., Zeh H.D., Kiefer C., Giulini D., Kupsch J., Stamatescu I.O.,
  Decoherence and the appearance of a classical world in quantum theory, 2nd
  ed., \href{http://dx.doi.org/10.1007/978-3-662-05328-7}{Springer-Verlag}, Berlin, 2003.

\bibitem{K78}
Kasuya M., The Einstein--Cartan theory of gravitation in a Hamiltonian form,
  \href{http://dx.doi.org/10.1143/PTP.60.167}{\textit{Progr. Theoret. Phys.}} \textbf{60} (1978), 167--177.

\bibitem{Kent}
Kent A., Might quantum-induced deviations from the {E}instein equations
  detectably af\/fect gravitational wave propagation?, \href{http://dx.doi.org/10.1007/s10701-013-9716-6}{\textit{Found. Phys.}}
  \textbf{43} (2013), 707--718, \href{http://arxiv.org/abs/1204.5961}{arXiv:1204.5961}.

\bibitem{Kouletsis}
Kouletsis I., Covariance and time regained in canonical general relativity,
  \href{http://dx.doi.org/10.1103/PhysRevD.78.064014}{\textit{Phys. Rev.~D}} \textbf{78} (2008), 064014, 22~pages,
  \href{http://arxiv.org/abs/0803.0125}{arXiv:0803.0125}.

\bibitem{Bubble}
Kucha{\v{r}} K.V., A bubble-time canonical formalism for geometrodynamics,
  \href{http://dx.doi.org/10.1063/1.1666050}{\textit{J.~Math. Phys.}} \textbf{13} (1972), 768--781.

\bibitem{Kuchar81b}
Kucha{\v{r}} K.V., General relativity: dynamics without symmetry,
  \href{http://dx.doi.org/10.1063/1.524842}{\textit{J.~Math. Phys.}} \textbf{22} (1981), 2640--2654.

\bibitem{Kuchar92}
Kucha{\v{r}} K.V., Time and interpretations of quantum gravity, in Proceedings
  of the 4th {C}anadian {C}onference on {G}eneral {R}elativity and
  {R}elativistic {A}strophysics ({W}innipeg, {MB}, 1991), Editors
  G.~Kunstatter, D.~Vincent, J.~Williams, World Sci. Publ., River Edge, NJ,
  1992, 211--314.

\bibitem{Kuchar93}
Kucha{\v{r}} K.V., Canonical quantum gravity, in General Relativity and
  Gravitation 1992 ({C}\'ordoba), Editors R.J.~Gleiser, C.N.~Kozamah, O.M.~Moreschi, Institute of Physics Publishing, Bristol, 1993, 119--150,
  \mbox{\href{http://arxiv.org/abs/gr-qc/9304012}{gr-qc/9304012}}.

\bibitem{Kuchar94}
Kucha{\v{r}} K.V., Geometrodynamics of {S}chwarzschild black holes,
  \href{http://dx.doi.org/10.1103/PhysRevD.50.3961}{\textit{Phys. Rev.~D}} \textbf{50} (1994), 3961--3981, \mbox{\href{http://arxiv.org/abs/gr-qc/9403003}{gr-qc/9403003}}.

\bibitem{K99}
Kucha\v{r} K.V., The problem of time in quantum geometrodynamics, in The
  Arguments of Time, Editor J.~Butterf\/ield, Oxford University Press, Oxford,
  1999, 169--196.

\bibitem{Lanczos}
Lanczos C., The variational principles of mechanics, University of Toronto
  Press, Toronto, Ont., 1949.

\bibitem{HallHaw6}
Langlois D., Hamiltonian formalism and gauge invariance for linear
  perturbations in inf\/lation, \href{http://dx.doi.org/10.1088/0264-9381/11/2/011}{\textit{Classical Quantum Gravity}} \textbf{11}
  (1994), 389--407.

\bibitem{LeeWald}
Lee J., Wald R.M., Local symmetries and constraints, \href{http://dx.doi.org/10.1063/1.528801}{\textit{J.~Math. Phys.}}
  \textbf{31} (1990), 725--743.

\bibitem{LR97}
Littlejohn R.G., Reinsch M., Gauge f\/ields in the separation of rotations and
  internal motions in the {$n$}-body problem, \href{http://dx.doi.org/10.1103/RevModPhys.69.213}{\textit{Rev. Modern Phys.}}
  \textbf{69} (1997), 213--275.

\bibitem{LP04}
Lusanna L., Pauri M., The physical role of gravitational and gauge degrees of
  freedom in general relativity. {II}.~{D}irac versus {B}ergmann observables
  and the objectivity of space-time, \href{http://dx.doi.org/10.1007/s10714-005-0218-5}{\textit{Gen. Relativity Gravitation}}
  \textbf{38} (2006), 229--267, \href{http://arxiv.org/abs/gr-qc/0407007}{gr-qc/0407007}.

\bibitem{Marolf09}
Marolf D., Solving the problem of time in mini-superspace: measurement of
  {D}irac observables, \href{http://dx.doi.org/10.1103/PhysRevD.79.084016}{\textit{Phys. Rev.~D}} \textbf{79} (2009), 084016,
  10~pages, \href{http://arxiv.org/abs/0902.1551}{arXiv:0902.1551}.

\bibitem{Bohm6}
Maudlin J., Quantum non-locality and relativity: metaphysical intimations of
  modern physics, Blackwell, Oxford, 2002.

\bibitem{BM}
Mukhanov V.F., Feldman H.A., Brandenberger R.H., Theory of cosmological
  perturbations, \href{http://dx.doi.org/10.1016/0370-1573(92)90044-Z}{\textit{Phys. Rep.}} \textbf{215} (1992), 203--333.

\bibitem{Nambu}
Nambu Y., Generalized {H}amiltonian dynamics, \href{http://dx.doi.org/10.1103/PhysRevD.7.2405}{\textit{Phys. Rev.~D}} \textbf{7}
  (1973), 2405--2412.


\bibitem{Page1}
Page D.N., Sensible quantum mechanics: are probabilities only in the mind?,
  \href{http://dx.doi.org/10.1142/S0218271896000370}{\textit{Internat.~J. Modern Phys.~D}} \textbf{5} (1996), 583--596,
  \href{http://arxiv.org/abs/gr-qc/9507024}{gr-qc/9507024}.

\bibitem{Page2}
Page D.N., Consciousness and the quantum, \href{http://arxiv.org/abs/1102.5339}{arXiv:1102.5339}.

\bibitem{PW83}
Page D.N., Wootters W.K., Evolution without evolution: dynamics described by
  stationary observables, \href{http://dx.doi.org/10.1103/PhysRevD.27.2885}{\textit{Phys. Rev.~D}} \textbf{27} (1983), 2885--2892.

\bibitem{PR}
Penrose R., Rindler W., Spinors and space-time. {V}ol.~1. Two-spinor calculus
  and relativistic f\/ields, \href{http://dx.doi.org/10.1017/CBO9780511564048}{\textit{Cambridge Monographs on Mathematical Physics}},
  Cambridge University Press, Cambridge, 1984.

\bibitem{PR+}
Penrose R., Rindler W., Spinors and space-time. {V}ol.~2. Spinor and twistor
  methods in space-time geometry, \href{http://dx.doi.org/10.1017/CBO9780511524486}{\textit{Cambridge Monographs on Mathematical Physics}},
  Cambridge University Press, Cambridge, 1986.

\bibitem{P78}
Pilati M., The canonical formulation of supergravity, \href{http://dx.doi.org/10.1016/0550-3213(78)90262-6}{\textit{Nuclear Phys.~B}}
  \textbf{132} (1978), 138--154.

\bibitem{Pitts}
Pitts J.B., A f\/irst class constraint generates not a gauge transformation, but
  a bad physical change: the case of electromagnetism, \href{http://arxiv.org/abs/1310.2756}{arXiv:1310.2756}.

\bibitem{Diffs1}
Polyakov A.M., Gauge transformations and dif\/feomorphisms, \href{http://dx.doi.org/10.1142/S0217751X90000386}{\textit{Internat.~J.
  Modern Phys.~A}} \textbf{5} (1990), 833--842.

\bibitem{PS05}
Pons J.M., Salisbury D.C., Issue of time in generally covariant theories and
  the {K}omar--{B}ergmann approach to observables in general relativity,
  \href{http://dx.doi.org/10.1103/PhysRevD.71.124012}{\textit{Phys. Rev.~D}} \textbf{71} (2005), 124012, 16~pages,
  \href{http://arxiv.org/abs/gr-qc/0503013}{gr-qc/0503013}.

\bibitem{PSS97}
Pons J.M., Salisbury D.C., Shepley L.C., Gauge transformations in the
  {L}agrangian and {H}amiltonian formalisms of generally covariant theories,
  \href{http://dx.doi.org/10.1103/PhysRevD.55.658}{\textit{Phys. Rev.~D}} \textbf{55} (1997), 658--668, \href{http://arxiv.org/abs/gr-qc/9612037}{gr-qc/9612037}.

\bibitem{PSS09b}
Pons J.M., Salisbury D.C., Sundermeyer K.A., Gravitational observables,
  intrinsic coordinates, and cano\-ni\-cal maps, \href{http://dx.doi.org/10.1142/S0217732309030473}{\textit{Modern Phys. Lett.~A}}
  \textbf{24} (2009), 725--732, \href{http://arxiv.org/abs/0902.0401}{arXiv:0902.0401}.

\bibitem{PSS09}
Pons J.M., Salisbury D.C., Sundermeyer K.A., Revisiting observables in
  generally covariant theories in the light of gauge f\/ixing methods,
  \href{http://dx.doi.org/10.1103/PhysRevD.80.084015}{\textit{Phys. Rev.~D}} \textbf{80} (2009), 084015, 23 pages,
  \href{http://arxiv.org/abs/0905.4564}{arXiv:0905.4564}.

\bibitem{PSS10}
Pons J.M., Salisbury D.C., Sundermeyer K.A., Observables in classical canonical
  gravity: folklore demystif\/ied, \href{http://dx.doi.org/10.1088/1742-6596/222/1/012018}{\textit{J.~Phys. Conf. Ser.}} \textbf{222}
  (2010), 012018, 15~pages, \href{http://arxiv.org/abs/1001.2726}{arXiv:1001.2726}.

\bibitem{Rov91a}
Rovelli C., Is there incompatibility between the ways time is treated in
  general relativity and in standard quantum mechanics?, in Conceptual Problems
  of Quantum Gravity ({N}orth {A}ndover, {MA}, 1988), \textit{Einstein Stud.},
  Vol.~2, Editors A.~Ashtekar, J.~Stachel, Birkh\"auser Boston, Boston, MA,
  1991, 126--140.

\bibitem{Rov91c}
Rovelli C., Quantum evolving constants. {R}eply to: ``{C}omment on: `{T}ime in
  quantum gravity: an hypothesis'\,'', \href{http://dx.doi.org/10.1103/PhysRevD.44.1339}{\textit{Phys. Rev.~D}} \textbf{44}
  (1991), 1339--1341.

\bibitem{Rov91b}
Rovelli C., Time in quantum gravity: an hypothesis, \href{http://dx.doi.org/10.1103/PhysRevD.43.442}{\textit{Phys. Rev.~D}}
  \textbf{43} (1991), 442--456.

\bibitem{Rov02a}
Rovelli C., G{PS} observables in general relativity, \href{http://dx.doi.org/10.1103/PhysRevD.65.044017}{\textit{Phys. Rev.~D}}
  \textbf{65} (2002), 044017, 6~pages, \href{http://arxiv.org/abs/gr-qc/0110003}{gr-qc/0110003}.

\bibitem{Rov02b}
Rovelli C., Partial observables, \href{http://dx.doi.org/10.1103/PhysRevD.65.124013}{\textit{Phys. Rev.~D}} \textbf{65} (2002),
  124013, 8~pages, \href{http://arxiv.org/abs/gr-qc/0110035}{gr-qc/0110035}.

\bibitem{Rovellibook}
Rovelli C., Quantum gravity, \href{http://dx.doi.org/10.1017/CBO9780511755804}{\textit{Cambridge Monographs on Mathematical Physics}},
  Cambridge University Press, Cambridge, 2004.

\bibitem{Rovelli13}
Rovelli C., Why gauge?, \href{http://dx.doi.org/10.1007/s10701-013-9768-7}{\textit{Found. Phys.}} \textbf{44} (2014), 91--104,
  \href{http://arxiv.org/abs/1308.5599}{arXiv:1308.5599}.

\bibitem{Bohm5}
Saunders S., The ``beables'' of relativistic pilot-wave theory, in From Physics
  to Philosophy ({C}ambridge, 1997), Editors M.~Redhead, J.~Butterf\/ield,
  C.~Pagonis, Cambridge University Press, Cambridge, 1999, 71--89.

\bibitem{Savvidou04}
Savvidou N., Histories approach to general relativity. {I}.~The spacetime
  character of the canonical description, \href{http://dx.doi.org/10.1088/0264-9381/21/2/020}{\textit{Classical Quantum Gravity}}
  \textbf{21} (2004), 615--630, \href{http://arxiv.org/abs/gr-qc/0306034}{gr-qc/0306034}.

\bibitem{Savvidou04b}
Savvidou N., Histories approach to general relativity. {II}.~{I}nvariance
  groups, \href{http://dx.doi.org/10.1088/0264-9381/21/2/021}{\textit{Classical Quantum Gravity}} \textbf{21} (2004), 631--646,
  \href{http://arxiv.org/abs/gr-qc/0306036}{gr-qc/0306036}.

\bibitem{Measurement1}
Schlosshauer M., Decoherence, the measurement problem, and interpretations of
  quantum mechanics, \href{http://dx.doi.org/10.1103/RevModPhys.76.1267}{\textit{Rev. Modern Phys.}} \textbf{76} (2005), 1267--1305.

\bibitem{HallHaw4}
Shirai I., Wada S., Cosmological perturbations and quantum f\/ields in curved
  space, \href{http://dx.doi.org/10.1016/0550-3213(88)90428-2}{\textit{Nuclear Phys.~B}} \textbf{303} (1988), 728--750.

\bibitem{Smolin}
Smolin L., The present moment in quantum cosmology: challenges to the arguments
  for the elimination of time, in Time and the Instant, Editor R.~Durie,
  Clinamen Press, Manchester, 2000, 112--143, \href{http://arxiv.org/abs/gr-qc/0104097}{gr-qc/0104097}.

\bibitem{Bohm10}
Smolin L., A real ensemble interpretation of quantum mechanics, \href{http://dx.doi.org/10.1007/s10701-012-9666-4}{\textit{Found.
  Phys.}} \textbf{42} (2012), 1239--1261, \href{http://arxiv.org/abs/1104.2822}{arXiv:1104.2822}.

\bibitem{Sni}
{\'S}niatycki J., Dirac brackets in geometric dynamics, \textit{Ann. Inst. H.~Poincar\'e Sect.~A} \textbf{20} (1974), 365--372.

\bibitem{Stewart}
Stewart J., Advanced general relativity, \textit{Cambridge Monographs on Mathematical
  Physics}, Cambridge University Press, Cambridge, 1990.

\bibitem{Bohm7}
Struyve W., Pilot-wave theory and quantum f\/ields, \href{http://dx.doi.org/10.1088/0034-4885/73/10/106001}{\textit{Rep. Progr. Phys.}}
  \textbf{73} (2010), 106001, 30~pages, \href{http://arxiv.org/abs/0707.3685}{arXiv:0707.3685}.

\bibitem{Szekeres}
Szekeres P., The gravitational compass, \href{http://dx.doi.org/10.1063/1.1704788}{\textit{J.~Math. Phys.}} \textbf{6}
  (1965), 1387--1391.

\bibitem{tHooft}
't~Hooft G., Determinism beneath quantum mechanics, in Quo Vadis Quantum
  Mechanics? (Philadelphia, 2002), Editors A.~Elitzur, S.~Dolev, N.~Kolenda,
  \href{http://dx.doi.org/10.1007/3-540-26669-0_8}{\textit{Frontiers Collection}}, Springer-Verlag, Berlin, 2005, 99--111,
  \href{http://arxiv.org/abs/quant-ph/0212095}{quant-ph/0212095}.

\bibitem{Takh}
Takhtajan L., On foundation of the generalized {N}ambu mechanics, \href{http://dx.doi.org/10.1007/BF02103278}{\textit{Comm.
  Math. Phys.}} \textbf{160} (1994), 295--315, \href{http://arxiv.org/abs/hep-th/9301111}{hep-th/9301111}.

\bibitem{Tambornino}
Tambornino J., Relational observables in gravity: a review, \href{http://dx.doi.org/10.3842/SIGMA.2012.017}{\textit{SIGMA}}
  \textbf{8} (2012), 017, 30~pages, \href{http://arxiv.org/abs/1109.0740}{arXiv:1109.0740}.

\bibitem{T73}
Teitelboim C., How commutators of constraints ref\/lect the spacetime structure,
  \href{http://dx.doi.org/10.1016/0003-4916(73)90096-1}{\textit{Ann. Physics}} \textbf{79} (1973), 542--557.

\bibitem{T77}
Teitelboim C., Supergravity and square roots of constraints, \href{http://dx.doi.org/10.1103/PhysRevLett.38.1106}{\textit{Phys. Rev.
  Lett.}} \textbf{38} (1977), 1106--1110.

\bibitem{Thiemannbook}
Thiemann T., Modern canonical quantum general relativity, \href{http://dx.doi.org/10.1017/CBO9780511755682}{\textit{Cambridge Monographs
  on Mathematical Physics}}, Cambridge University Press, Cambridge, 2007.

\bibitem{Torre93}
Torre C.G., Gravitational observables and local symmetries, \href{http://dx.doi.org/10.1103/PhysRevD.48.R2373}{\textit{Phys.
  Rev.~D}} \textbf{48} (1993), R2373--R2376, \mbox{\href{http://arxiv.org/abs/gr-qc/9306030}{gr-qc/9306030}}.

\bibitem{TorreGowdy}
Torre C.G., Observables for the polarized {G}owdy model, \href{http://dx.doi.org/10.1088/0264-9381/23/5/007}{\textit{Classical
  Quantum Gravity}} \textbf{23} (2006), 1543--1555, \href{http://arxiv.org/abs/gr-qc/0508008}{gr-qc/0508008}.

\bibitem{+Torre1}
Torre C.G., Anderson I.M., Symmetries of the {E}instein equations,
  \href{http://dx.doi.org/10.1103/PhysRevLett.70.3525}{\textit{Phys. Rev. Lett.}} \textbf{70} (1993), 3525--3529,
  \href{http://arxiv.org/abs/gr-qc/9302033}{gr-qc/9302033}.

\bibitem{NSI2}
Unruh W.G., Wald R.M., Time and the interpretation of canonical quantum
  gravity, \href{http://dx.doi.org/10.1103/PhysRevD.40.2598}{\textit{Phys. Rev.~D}} \textbf{40} (1989), 2598--2614.

\bibitem{Algebroid2}
Vaisman I., Lectures on the geometry of {P}oisson manifolds, \href{http://dx.doi.org/10.1007/978-3-0348-8495-2}{\textit{Progress
  in Mathematics}}, Vol.~118, Birkh\"auser Verlag, Basel, 1994.

\bibitem{VM10}
Vargas~Moniz P., Quantum Cosmology~-- the supersymmetric perspective. Vol.~1.
  Fundamentals, \href{http://dx.doi.org/10.1007/978-3-642-11575-2}{\textit{Lecture Notes in Phys.}}, Vol.~803, Springer, Berlin,
  2010.

\bibitem{Bohm1}
Vink J.C., Quantum mechanics in terms of discrete beables, \href{http://dx.doi.org/10.1103/PhysRevA.48.1808}{\textit{Phys.
  Rev.~A}} \textbf{48} (1993), 1808--1818.

\bibitem{HallHaw3}
Wada S., Consistency of canonical quantization of gravity and boundary
  conditions for the wave function of the {U}niverse, \href{http://dx.doi.org/10.1103/PhysRevD.34.2272}{\textit{Phys. Rev.~D}}
  \textbf{34} (1986), 2272--2276.

\bibitem{HallHaw2}
Wada S., Quantum cosmological perturbations in pure gravity, \href{http://dx.doi.org/10.1016/0550-3213(86)90073-8}{\textit{Nuclear
  Phys.~B}} \textbf{276} (1986), 729--743, Erratum, \href{http://dx.doi.org/10.1016/0550-3213(87)90060-5}{\textit{Nuclear
  Phys.~B}} \textbf{284} (1987), 747--748.

\bibitem{Measurement2}
Wallace D., Philosophy of quantum mechanics, in The {A}shgate Companion to
  Contemporary Philosophy of Physics, Editor D.~Rickles, Ashgate, Aldershot,
  2008, 16--98, \href{http://arxiv.org/abs/0712.0149}{arXiv:0712.0149}.

\bibitem{Weinberg}
Weinberg S., The quantum theory of f\/ields. {V}ol.~{I}. Foundations, Cambridge
  University Press, Cambridge, 2005.

\bibitem{Weinberg+}
Weinberg S., The quantum theory of f\/ields. {V}ol.~{II}. Modern applications,
  Cambridge University Press, Cambridge, 2005.

\bibitem{Battelle}
Wheeler J.A., Superspace and the nature of quantum geometrodynamics, in
  Battelle Rencontres: 1967 Lectures in Mathematics and Physics, Editors
  C.~DeWitt, J.A.~Wheeler, Benjamin, New York, 1968, 242--307.

\bibitem{WuthrichTh}
W\"{u}thrich C., Approaching the Planck scale from a generally relativistic
  point of view: a philosophical appraisal of loop quantum gravity, Ph.D.\   Thesis, University of Pittsburgh, 2006.

\end{thebibliography}
\end{document}